\documentclass[useAMS,usegraphicx]{mn2e}
\usepackage{amssymb}
\usepackage{graphicx}
\usepackage{times}
\usepackage{amssymb}
\usepackage{epstopdf}
\usepackage{graphicx}
\usepackage{footnote}
\usepackage{threeparttable}
\usepackage{float}
\usepackage{pdflscape}
\usepackage[normalem]{ulem}

\title[Near-Infrared and optical studies of Nova Aquilae 2015]{Near-infrared and optical studies of  the highly obscured nova V1831 Aquilae (Nova Aquilae 2015)}

\author[Banerjee et al.]{D. P. K. Banerjee$^{1}$\thanks{E-mail: orion@prl.res.in}, Mudit K. Srivastava$^{1}$, N. M. Ashok$^{1}$,  U.Munari$^{2}$, F.-J.Hambsch$^{3}$,
\and G.L.Righetti$^{3}$  $\&$ A.Maitan$^{3}$\\
$^{1}$Astronomy and Astrophysics Division, Physical Research Laboratory, Navrangpura, Ahmedabad, India, 380009 \\
$^{2}$INAF Astronomical Observatory of Padova, via dell'Osservatorio 8, 36012 Asiago (VI), Italy\\
$^{3}$ANS Collaboration, c/o Astronomical Observatory, 36012 Asiago (VI), Italy
}

\begin{document}

\date{Accepted YYYY Month DD.  Received YYYY Month DD; in original form YYYY Month DD}

\maketitle

\label{firstpage}

\begin{abstract}

Near Infrared (NIR) and optical photometry and spectroscopy are presented for the  nova V1831 Aquilae, covering the early decline and dust forming phases during the first $\sim$90 days after its discovery. The nova is highly reddened due to interstellar extinction. Based solely on the nature of NIR spectrum we are able to classify the nova to be of the Fe II class. The distance and extinction to the nova are estimated to be  6.1 $\pm$ 0.5 kpc and $A_{\rm v}$ $\sim$ 9.02 respectively. Lower limits of the electron density, emission measure and ionized ejecta mass are made from a Case B analysis of the NIR Brackett lines while the neutral gas mass is estimated from the optical [OI] lines. We discuss the cause for a rapid strengthening of the He I 1.0830 $\mu$m  line during the early stages. V1831 Aql formed a modest amount of dust fairly early ($\sim$ 19.2 days after discovery); the dust shell is not seen to be  optically thick. Estimates are made of the dust temperature, dust mass and grain size. Dust formation  commences around day 19.2  at a condensation temperature of 1461 $\pm$ 15 K, suggestive of a carbon composition, following which the  temperature is seen to gradually decrease to 950K. The dust mass shows a rapid initial increase which we interpret as being due to  an increase in the number of grains,  followed by a period of constancy suggesting the absence of grain destruction processes during this latter time. A discussion is made of the evolution of these  parameters,  including certain peculiarities  seen in the grain radius evolution.


\end{abstract}


\section{Introduction}
 V1831 Aql was first detected  as a new transient source by the All Sky Automated Survey for Supernovae and designated as ASASSN-15qd by Shappee et al. (2015) on 2015 September 29.31 UT at $V$ $\sim$ 14.7 (error 0.1 mag) under hazy sky conditions. A more robust confirmation was subsequently made at $V$ = 15.2 on 2015 October 1.29 UT. An archival search of the Kamogata-Kiso-Kyoto Wide Field Survey (Shappee et al. 2015 and references within) showed a pre-discovery detection of the source at $I_c$ = 10.5 on 2015 September 27.409 UT. Shappee et al. (2015) suggested that the new transient was likely to be a highly obscured galactic nova with the large $E(V-I_c)$ value of $\sim$  4.6 suggesting a considerable interstellar extinction. The search for the quiescent counterpart of the nova in the UKIDSS Galactic Plane Survey data base by Maccarone (2015) resulted in the detection of a source with $K$ =  17.1 $\pm$ 0.1 (UGPS J192150.15+150924.8) with good positional coincidence. Assuming this source to be the nova progenitor  and by adopting  the required estimates involving the extinction to the nova, the intrinsic colors of cataclysmic variables at quiescence, and the distance to the nova, Maccarone (2015) infer that the absolute magnitude of this potential progenitor should be $M_v$  $\sim$ 3.8 making it easily faint enough to have been undetected in pre-outburst optical data. We discuss this further in section 3.1 when estimating the outburst amplitude of the nova. Close to the Shappee et al.(2015) discovery, V1831 Aql was also detected by Itagaki (Nakano 2015) at an unfiltered magnitude of 12.4 on 2015 October 5.548 UT; the source was designated as PNV J19215012+1509248 = Nova Aquilae 2015.  Spectroscopic confirmation of the source as a nova was obtained on 2015 October 5.598 and 6.482 UT by Maehara $\&$  Fujii (2015) and Fujii(2015) who noted the strong H$\alpha$ emission (FWHM = 1600 km/s) in the spectrum and the presence of emission lines of Fe II (multiplets 73,74), O I, Ca II and [O I]. However, the authors did  not attempt to classify whether the nova was of the Fe II or He/N class. This may be due to considerable faintness in the blue part of the spectrum  because of large reddening  i.e. very low signal in the blue region where the strongest of the Fe II lines - generally used to identify the Fe II novae - are  expected to register. Because NIR region is more amenable for observations in directions of high extinction, NIR spectra in the 0.9 to 2.5 micron region were less affected and these clearly established the nova to be of the Fe II class (Ashok et al. 2015). The Fe II  classification was later confirmed by Goranskij $\&$ Barsukova (2015) in the optical.
\par
In this paper we present our  NIR and optical spectroscopic and photometric observations of V1831 Aql, preliminary reports of which were made in Ashok, Banerjee $\&$ Srivastava (2015) and Srivastava, Banerjee $\&$ Ashok (2015).


\section{Observations}
\label{sec-Obs}

\begin{table}
\centering
\caption{Log of the optical photometry}
\begin{tabular}{ccccccc}
\hline
Date (UT)      & Time$^a$     &   R$_{c}$  &   I$_{c}$  &   Tel.$^b$        \\
(yyyy-mm-dd.dd)   & (days)       &  (mags)    &  (mags)   &           \\
               &      &            &             &          \\
\hline
\hline
2015-10-08.82    &  9.32 & 13.300 $\pm$0.016  & 11.240 $\pm$0.013   & 220  \\
 2015-10-11.80    & 12.30 & 13.404 $\pm$0.018  & 11.344 $\pm$0.019   & 220  \\
 2015-10-11.88    & 12.38 & 13.469 $\pm$0.011  & 11.404 $\pm$0.006   & 157  \\
 2015-10-12.07    & 12.56 &                    & 11.413 $\pm$0.010   & 210  \\
 2015-10-13.07    & 13.57 &                    & 11.485 $\pm$0.010   & 210  \\
 2015-10-14.07    & 14.57 &                    & 11.374 $\pm$0.013   & 210  \\
 2015-10-15.07    & 15.57 &                    & 11.456 $\pm$0.011   & 210  \\
 2015-10-16.07    & 16.57 &                    & 11.653 $\pm$0.015   & 210  \\
 2015-10-19.78    & 20.28 & 14.005 $\pm$0.012  & 11.966 $\pm$0.006   & 157  \\
 2015-10-20.81    & 21.31 & 14.001 $\pm$0.011  & 11.895 $\pm$0.005   & 157  \\
 2015-10-21.05    & 21.55 &                    & 12.021 $\pm$0.019   & 210  \\
 2015-10-21.74    & 22.24 & 14.124 $\pm$0.011  & 11.981 $\pm$0.005   & 157  \\
 2015-10-22.05    & 22.55 &                    & 11.981 $\pm$0.021   & 210  \\
 2015-10-23.05    & 23.55 &                    & 12.091 $\pm$0.022   & 210  \\
 2015-10-23.84    & 24.34 & 14.317 $\pm$0.015  & 12.216 $\pm$0.010   & 157  \\
 2015-10-24.75    & 25.25 & 14.367 $\pm$0.016  & 12.289 $\pm$0.010   & 157  \\
 2015-10-30.74    & 31.24 & 14.799 $\pm$0.020  & 12.757 $\pm$0.010   & 157  \\
 2015-10-31.75    & 32.25 & 14.772 $\pm$0.024  & 12.828 $\pm$0.027   & 220  \\
 2015-10-31.77    & 32.27 & 14.892 $\pm$0.014  & 12.822 $\pm$0.010   & 157  \\
 2015-11-01.79    & 33.29 & 14.858 $\pm$0.020  & 12.912 $\pm$0.021   & 220  \\
 2015-11-02.73    & 34.23 & 15.152 $\pm$0.019  & 13.030 $\pm$0.012   & 157  \\
 2015-11-02.79    & 34.29 & 15.080 $\pm$0.025  & 13.128 $\pm$0.018   & 220  \\
 2015-11-06.73    & 38.23 & 15.480 $\pm$0.027  & 13.417 $\pm$0.011   & 157  \\
 2015-11-08.73    & 40.23 & 15.646 $\pm$0.016  & 13.589 $\pm$0.010   & 157  \\
 2015-11-08.75    & 40.25 & 15.452 $\pm$0.015  & 13.510 $\pm$0.012   & 220  \\
 2015-11-12.00    & 43.50 &                    & 13.761 $\pm$0.047   & 210  \\
 2015-11-13.01    & 44.51 &                    & 13.952 $\pm$0.055   & 210  \\
 2015-11-15.01    & 46.51 &                    & 14.173 $\pm$0.300   & 210  \\
 2015-11-16.01    & 47.51 &                    & 13.742 $\pm$0.043   & 210  \\
 2015-11-17.01    & 48.51 &                    & 13.999 $\pm$0.065   & 210  \\
 2015-11-18.01    & 49.51 &                    & 14.184 $\pm$0.062   & 210  \\
 2015-11-20.01    & 51.51 &                    & 13.932 $\pm$0.054   & 210  \\
 2015-11-21.01    & 52.51 &                    & 14.089 $\pm$0.073   & 210  \\
 2015-11-22.72    & 54.22 & 16.513 $\pm$0.049  & 14.790 $\pm$0.029   & 220  \\
 2015-11-27.73    & 59.23 & 16.554 $\pm$0.041  & 14.882 $\pm$0.019   & 220  \\
 2015-12-10.75    & 72.25 & 16.650 $\pm$0.030  & 15.732 $\pm$0.032   & 140  \\
 2015-12-21.69    & 83.19 & 17.110 $\pm$0.032  & 16.290 $\pm$0.030   & 140  \\
\hline
\hline
\end{tabular}
\label{table-OptPhot}

\begin{list}{}{}
 \item a : Measured from  JD2457295.00 (2015-09-29.50) = t$_{0}$
 \item b : Telescope number of the ANS consortium

 \end{list}
\end{table}

\begin{table}
\centering
\caption{Log of the optical spectroscopic observations  from Asiago}
\begin{tabular}{cccrccc}
\hline

Date (UT)      & Time     &    Exp. & Range &\AA/pix    & Instrument        \\
(yyyy-mm-dd)   & (days)        &    time & (nm)     &  & $\&$      \\
               &   &    (s)  &         &  &Telescope       \\
\hline
\hline

2015-10-16.80 & 17.30 &  240 & 330-800 &   2.31   & B\&C, 1.22m \\
2015-10-20.84 & 21.34 &  480 & 330-800    & 2.31   & B\&C,  1.22m \\
2015-10-23.81 & 24.31 &  600 & 330-800    & 2.31   & B\&C, 1.22m \\
2015-10-23.85 & 24.35 & 1800 & 370-740    &  & Echelle,  1.82m \\
 &  &  &     &  &R = 20,000\\
2015-11-03.77 & 35.28 & 1200 & 330-800    & 2.31   & B\&C, 1.22m \\
2015-11-06.72 & 38.22 & 1800 & 330-800    & 2.31   & B\&C,  1.22m \\
2015-11-24.74 & 56.24 &  900 & 330-800    & 2.31   & B\&C,  1.22m \\
2015-12-10.71 & 72.21 & 1800 & 624-696    & 0.39   & AFOSC, 1.82m \\
2015-12-10.74 & 72.24 & 1800 & 560-1020   & 2.95   & AFOSC, 1.82m \\

\hline
\hline
\end{tabular}
\label{table-OptSpec}

\end{table}


\begin{table}
\centering
\caption{Log of the NIR spectroscopy from Mt. Abu}
\begin{tabular}{lcccccc}
\hline
Date (UT)          & Time                   &&Exposure  time &   Std.             \\
(yyyy-mm-dd.dd)      & (days)                      && (s)&             star\\
             &                              && $(IJ, JH, HK)$  &                   \\
\hline
\hline

             & & \\
 2015-10-08.75 &   9.25  &&  ( 300, 240,  240) &   SAO 104779 \\
 2015-10-16.72 &  17.22  &&  ( 600, 480,  480) &   SAO 104779 \\
 2015-10-17.68 &  18.18  &&  ( 600, 600,  600) &   SAO 104779 \\
 2015-10-18.71 &  19.21  &&  ( 600, 600,  600) &   SAO 104779 \\
 2015-10-19.72 &  20.22  &&  ( 600, 600,  600) &   SAO 104779 \\
 2015-10-26.65 &  27.15  &&  ( ---, 540,  600) &   SAO 104779 \\
 2015-10-27.67 &  28.17  &&  ( 600, 600,  600) &   SAO 104779 \\
 2015-11-04.68 &  36.18  &&  ( 300, 360,  360) &   SAO 104779 \\
 2015-11-05.60 &  37.10  &&  ( 600, 600,  600) &   SAO 104779 \\
 2015-11-19.63 &  51.13  &&  ( 380, 760,  760) &   SAO 104779 \\
 2015-11-26.61 &  58.11  &&  ( 380, 760,  760) &   SAO 104804 \\
 2015-11-27.62 &  59.12  &&  ( ---, 380,  760) &   SAO 104804 \\
 2015-12-02.57 &  64.07  &&  ( 380, 760,  760) &   SAO 104804 \\
 2015-12-18.56 &  80.06  &&  ( 380, 380,  760) &   SAO 104804 \\
\hline
\hline
\end{tabular}
\label{table-NIRSpec}

\begin{list}{}{}
\item
\end{list}
\end{table}


\begin{table}
\centering
\caption{Log of the near-IR photometry from Mt. Abu}
\begin{tabular}{ccccccc}
\hline
Date (UT)      & Time     &   $J$  &   $H$   &  $K$$_{s}$ &        \\
(yyyy-mm-dd.dd)   & (days)       &  (mags)    &  (mags)     &  (mags)        \\
               &      &            &             &          \\
\hline
\hline

2015-10-08.79	&  9.29 & 8.32	$\pm$  0.05 &	7.75   $\pm$ 0.03 & 7.09 $\pm$	0.04	\\
2015-10-16.76	& 17.26	& 8.92	$\pm$  0.06 &	8.18   $\pm$ 0.04 & 7.12 $\pm$	0.07	\\
2015-10-17.72	& 18.22	& 8.90	$\pm$  0.07 &	8.12   $\pm$ 0.11 & 6.91 $\pm$	0.06	\\
2015-10-18.75	& 19.25	& 8.97	$\pm$  0.08 &	8.13   $\pm$ 0.02 & 6.89 $\pm$	0.10	\\
2015-10-19.76	& 20.26	& 9.13	$\pm$  0.04 &	8.02   $\pm$ 0.05 & --	        --      \\
2015-10-26.69   & 27.19 & 9.18  $\pm$  0.06 &   7.38   $\pm$ 0.05 & 5.76 $\pm$  0.05    \\
2015-10-27.71   & 28.21 & 9.15  $\pm$  0.08 &   7.32   $\pm$ 0.03 & 5.71 $\pm$  0.05    \\
2015-11-04.72	& 36.22	& 9.29	$\pm$  0.04 &	7.20   $\pm$ 0.04 & 5.26 $\pm$	0.11	\\
2015-11-05.64	& 37.14	& 9.40	$\pm$  0.04 &	7.33   $\pm$ 0.06 & 5.57 $\pm$	0.05	\\
2015-11-20.63	& 52.13	& 10.36	$\pm$  0.06 &	8.08   $\pm$ 0.04 & 6.14 $\pm$	0.05	\\
2015-12-01.58	& 63.08	& 11.10	$\pm$  0.04 &	8.73   $\pm$ 0.02 & 6.69 $\pm$	0.02	\\
2015-12-17.56	& 79.06	& 11.83	$\pm$  0.07 &	9.65   $\pm$ 0.10 & 7.46 $\pm$	0.07	\\
2015-12-19.56	& 81.06	& 11.91	$\pm$  0.03 &	9.74   $\pm$ 0.02 & 7.61 $\pm$	0.08	\\
2015-12-25.54 	& 87.04	& 12.20	$\pm$  0.05 &	--	   --     & 8.00 $\pm$	0.05	\\
\hline
\hline
\end{tabular}
\label{table-NIRPhot}

\end{table}

\begin{figure}
\centering
\includegraphics[bb=56 172 453 468,width= 3.0in,clip]{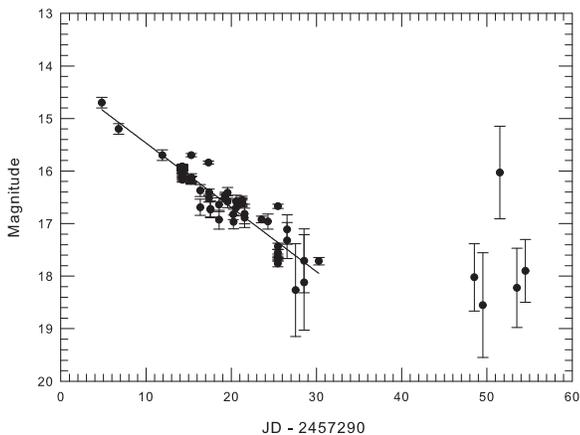}
\caption[]{The $V$ band light curve from AAVSO data. A  linear fit to the segment of the observed data, which was used to
estimate $t$$_{2}$, is shown. Additional details about the light-curve are discussed in the text.}
\label{fig-LC-AAVSO}
\end{figure}

\begin{figure}
\centering
\includegraphics[angle=0,width=0.5\textwidth]{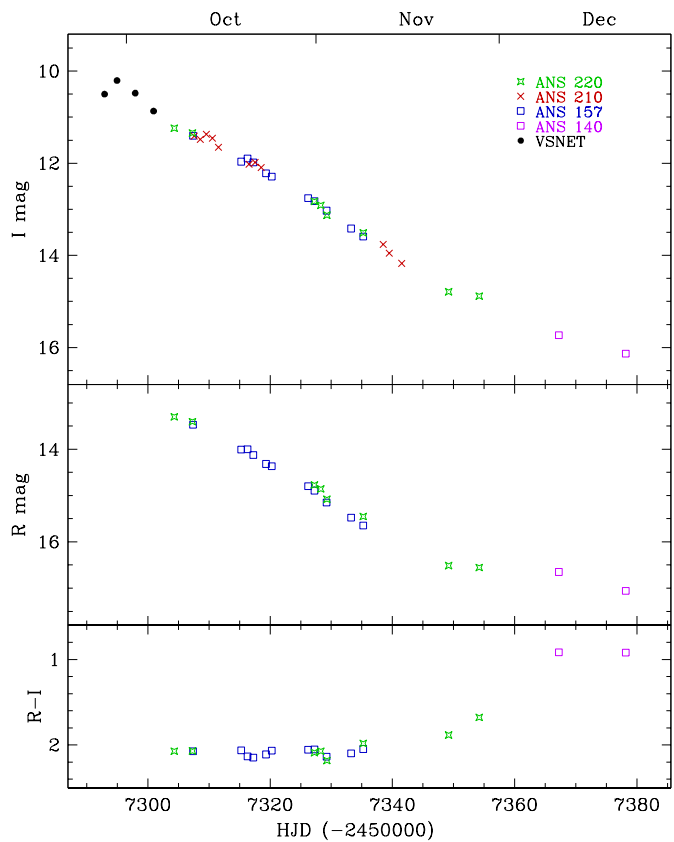}
\caption[]{ The lightcurves in the R$_{c}$ and I$_{c}$ bands  from data collected by different telescopes (shown by different symbols) of  the ANS consortia, Italy. The  halt of the decline in the  R$_{c}$ band around JD 7350 is due to the emerging dominance of Balmer H$\alpha$, that declines slower than the continuum.}
\label{fig-LC-RI}
\end{figure}

\begin{figure}
\centering
\includegraphics[bb=105 67 519 367, width= 3.0in,clip]{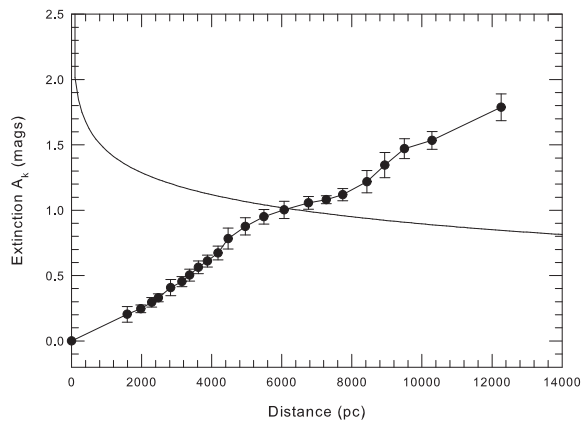}
\caption[]{ The curve joining the  data points (black circles) shows the variation of the  extinction towards V1831 Aquilae based on results from Marshall et al. (2006). The continuous straight line is a plot of extinction $A$ versus distance $d$ from the equation m$_{\rm v}$ - M$_{\rm v}$ = 5logd - 5 + A$_{\rm v}$
  where  m$_{\rm v}$ is known from observations and M$_{\rm v}$ is estimated from MMRD relations. The intersection of the two curves permits a simultaneous estimation of the extinction and distance to the nova. More details are given in the text.}
\label{fig-MMRD}
\end{figure}

\subsection{Optical photometry and spectroscopy observations}
 Optical photometry of Nova Aql 2015 was obtained with ANS Collaboration telescopes N. 140,157, 210 and 220. The nova was observable only in R$_{c}$ and I$_{c}$ bands because of its faintness and extremely red colors. The same local photometric sequence, spanning a wide color range and carefully calibrated against Landolt (2009) equatorial standards, was used at all telescopes and observing epochs, ensuing a high consistency among different data sets. Table~\ref{table-OptPhot} gives the details of R$_{C}$ and I$_{C}$ band photometry, where the quoted uncertainties are the total error budget, which quadratically combines the measurement error on the variable with the error associated to the transformation from the local to the standard photometric system (as defined by the photometric comparison sequence around the nova linked to the Landolt's standards).  The operation of ANS Collaboration telescopes is described in detail by Munari et al. (2012) and Munari \& Moretti (2012).  They are all located in Italy.  All measurements on the program nova were carried out with aperture photometry.Colors and magnitudes are obtained separately during the reduction process, and are not derived one from the other.
\par
Low resolution 3300-8000 $\AA$ spectra of V1831 Aql were obtained with the 1.22m telescope + B\&C spectrograph operated in Asiago by the Department of Physics and Astronomy of the University of Padova.  The CCD camera is an ANDOR iDus DU440A with a back-illuminated E2V 42-10 sensor, 2048$\times$512 array of 13.5 $\mu$m pixels.  A low- and a medium-resolution 5600-10200 \AA\  spectra of V1831 Aql were obtained with AFOSC spectrograph and imager mounted on the 1.82m telescope operated in Asiago by INAF Astronomical Observatory of Padova.  The CCD camera is an Andor DW436-BV which houses an E2V CCD42-40 AIMO back-illuminated CCD as detector, and the dispersing elements are VPH grisms. Finally, a high resolution spectrum of V1831 Aql was recorded at a resolving power of 20,000 with the Echelle spectrograph mounted on the Asiago 1.82m telescope. Table~\ref{table-OptSpec} provides a log of the spectroscopic observations.

\subsection{Near-Infrared observations from Mount Abu}

Nova Aql 2015 was observed from the 1.2m  optical/NIR telescope of the Physical Research Laboratory at Mount Abu Infrared observatory, India  using the Near-Infrared Camera and Spectrometer (NICS). NICS contains a re-imaging optical chain along with a 1024$\times$1024 HgCdTe Hawaii array as detector. It provides a resolution of $\sim$ 1000 in spectroscopy mode and covers $J$, $H$ and $K_s$ wavelength bands from 0.85-2.4 $\mu$m. Three different orientations of NICS gratings are used to record spectra in $IJ$, $JH$ and $HK$ regions with suitable overlaps. Photometry and spectroscopy of Nova Aql 2015 were done using NICS for a total 19 epochs from October 8, 2015 to December 25, 2016. For telluric line corrections in the spectra of Nova Aql 2015, spectra of a standard star (SAO 104779/FI V spectral class and SAO 104804/A0 spectral class; details are given in Table~\ref{table-NIRSpec}) were also taken at similar airmass. All the spectra were obtained in two dithered positions of the objects along the slit. Dark and sky lines corrected spectral frames were obtained by subtracting one from another. The data was later reduced using self developed image processing codes in C with {\it fitsio} library functions and using IRAF tasks. Wavelength calibrations were applied by identifying OH sky lines and telluric lines. Several sets of reduced spectra were then co-added. The standard star spectra showed the presence of hydrogen Paschen and Brackett lines. These lines were identified and removed. Nova spectra were then ratioed with these cleaned standard star spectra and further multiplied by the blackbody distribution function corresponding to the effective temperature of the standard star. Aperture photometry of the Nova Aql 2015 in {\it J, H} and {\it K$_{s}$} was performed along with photometric standard stars to estimate magnitudes following standard procedures (e.g. Banerjee \& Ashok 2002).  The details of the NIR  observations are given in  Tables~\ref{table-NIRSpec} and ~\ref{table-NIRPhot}.


\section{Results}
\label{sec-NAql15-results}

\subsection{Light curve, extinction, distance and outburst amplitude}
\label{subsec-NAql15-LCExt}

The $V$ band lightcurve from AAVSO data is presented in Figure~\ref{fig-LC-AAVSO} while the R$_{c}$ and I$_{c}$ band lightcurves are shown in Figure~\ref{fig-LC-RI}. The absolute magnitude M$_{\rm v}$  of the nova is being estimated from  maximum magnitude versus rate of decline (MMRD) relations given by della Valle \& Livio (1995) and Downes \& Duerbeck (2000). These relations correlate the absolute magnitude M$_{\rm v}$  with the time to decline by 2 magnitudes from visual maximum (t$_{2}$) which is a measure of the speed class of a nova. To estimate t$_{2}$, four discrepant data points had to be excluded from the AAVSO lightcurve. These rejected points were isolated outliers and obviously erroneous, lying as much as $\sim$ 2 to 2.5 mags away from their neighboring data. Figure~\ref{fig-LC-AAVSO} also includes a few reported magnitudes from Shappee et al (2015) and Nakano (2015). A least-squares minimised linear fit to the data  gives t$_{2}$ = 16.4 $\pm$ 0.5d assuming 29.5 September 2015 as time of optical maximum as inferred from the I$_{\rm c}$ lightcurve of Figure~\ref{fig-LC-RI}. In all subsequent discussion, we take 2015 September 29.5 UT as the reference time (t$_{0}$).  t$_{2}$ = 16.4 $\pm$ 0.5d puts this nova in the fast speed class (Payne-Gaposchkiin, 1957 and references therein). We discuss at a later stage the expected time scale for dust formation in novae of this speed class (Williams et al. 2013). Adopting m$_{\rm v}$(max) = 14.7 (Shappee et al. 2015) yields a value of the absolute magnitude of M$_{\rm v}$ = -8.29 $\pm$ 0.20  from the MMRD relation of della Valle \& Livio (1995). The MMRD relation of Downes \& Duerbeck (2000) gives a consistent value of M$_{\rm v}$ = -8.22 $\pm$ 0.83.
\par
We estimate the extinction and distance to the nova in the following way. If a reliable extinction versus distance plot is available in the nova's direction (e.g as the Marshall et al. (2006) data shown in Figure~\ref{fig-MMRD}) and a valid MMRD relation for the nova is also available, then estimates of both the extinction and distance can be simultaneously made. The continuous curve in Figure~\ref{fig-MMRD} shows the A$_{\rm k}$ versus distance curve from the relation  m$_{\rm v}$(max) - M$_{\rm v}$ = 5logd - 5 + A$_{\rm v}$  wherein we use m$_{v}$(max) = 14.7, M$_{\rm v}$ = -8.3 and convert A$_{\rm v}$ to A$_{\rm k}$ using A$_{\rm k}$ = 0.112$\times$ A$_{\rm v}$ (Rieke $\&$ Lebofsky, 1985). The second curve in Figure~\ref{fig-MMRD} is the extinction versus distance plot from  Marshall et al. (2006) based on modeling of the galactic extinction. The intersection of both curves should give the nova's distance and extinction because both curves are in principle making use of the same extinction. In the present case, we obtain $d$ = 6.1 $\pm$ 0.5 kpc and A$_{\rm k}$ = 1.01 $\pm$ 0.03. Using A$_{\rm k}$/A$_{v}$ = 0.112 and A$_{\rm v}$ = 3.09$E(B-V)$ (Rieke $\&$ Lebofsky, 1985) we get A$_{\rm v}$ = 9.02 and $E(B-V)$ = 2.92. We adopt this value of reddening in future calculations. Shappee et al. (2015) derive a larger value of Av = 11.5 compared to $A_{v}$ = 9.02 that we estimate. Possible ways of reconciling this difference may be as follows. They assume the nova's intrinsic colors at around maximum to be $(B-V)$ = $(V-I)$ = 0 whereas novae at maximum are statistically expected to have an intrinsic $(B-V)$ color  at maximum = 0.23 $\pm$ 0.06 with a 1$\sigma$ dispersion of 0.16 magnitudes (van den Bergh $\&$ Younger, 1987). Indeed novae do show colors between A to F spectral types at maximum (Gehrz, 1988). For e.g. V1500 Cyg at maximum showed a T(eff) = 6670K (Gehrz 1988) corresponding to F4-F5 spectral type which have intrinsic $V-I$ color of $\sim$ 0.5. Thus their estimate that $E(V-I)$ = 4.6 around maximum could have substantial uncertainty (they derive this using $V$ = 15.2 $\pm$ 0.1 is from 2015 Oct 01.29 and I$_{c}$ =10.6 $\pm$ 0.1 from  a day later on 2015 Oct 02.4356. Note the large errors on $V$ and I$_{c}$). Thus  $E(V-I_{c})$ = 4.6 could be substantially overestimated leading to inflated $A_{v}$ values via their  use of the equation $A_{v}/E(V-I_{C})$ = 2.49 (Stanek 1996). On the other hand, a lacuna in the approach we have adopted, is that the Marshall et al.(2006) extinction curve that we have used is averaged over a 5 arc minute square field and not exactly in the direction of the nova. But extinction can be very patchy and vary considerably over this field; thus the possibility that we have underestimated $A_{v}$ cannot be ruled out.
\par
One may like to compare the Marshall et al. (2006) extinction data in Figure~\ref{fig-MMRD} with other independent estimates. The Marshall et al. (2006) plot in Figure~\ref{fig-MMRD} shows that the extinction in the direction of the nova, up to the maximum distance d = 12.26 kpc for which it could be modeled, is A$_{\rm k}$ = 1.79 $\pm$ 0.10. In comparison, Schlafly $\&$ Finkbeiner (2011) give a mean value of A$_{\rm k}$ $\sim$ 2.09 along the entire line of sight in the nova's direction. So the agreement is reasonable.
\par
Novae, considering  all speed classes, are found to have a mean absolute magnitude in  quiescence  between 3.4 to 4.2 magnitudes;  fast novae in particular are expected to have a mean absolute magnitude of 3.7 (Warner, 1995).  Adopting this  mean quiescent value of M$_{\rm v}$ = 3.7 as reasonably representative for  nova V1831 Aql, and using   $d$ = 6.1 $\pm$ 0.5 kpc and  A$_{\rm v}$ = 9.02 as determined earlier,   the quiescent apparent magnitude of the nova is expected  to be 26.6 $\pm$ 0.2 magnitudes.  At this level of faintness, the progenitor is not expected to be seen in the optical archival images (also refer Maccarone 2015). Since maximum brightness was at m$_{\rm v}$ = 14.7 (Shappee et al. 2015), the outburst amplitude is estimated to be $\sim$ 12 magnitudes. That places it at a conventional position, frequented by  other classical novae (CNe), in the amplitude of outburst versus t$_{\rm 2}$ plot of CNe by Warner (1995, Figure 5.4).
\par
A few comments may be made regarding the evolution of the R$_{c}$ -I$_{c}$  colors shown in the bottom panel of Figure 2. It must be noted that both the R$_{c}$ and I$_{c}$ band continuum fluxes are greatly affected by the presence of very strong lines covered by the respective filters. There is the H$\alpha$ line in the R$_{c}$ band and similarly the OI 7774, 8446$\AA$ lines in the I$_{c}$ band. Although the OI 8446$\AA$ line is not covered in most of our spectra (except for the 2015 December spectrum) and we can only speculate about its exact contribution to  the continuum on other days, it is certainly expected to be stronger than OI 7774$\AA$ (which is already strong) because of  additional contribution from the Ly$\beta$ fluorescence presence. OI 8446$\AA$ is a strong line in novae and the OI 8446/7774 line ratio in Fe II novae can easily extend from unity to a  value of few times ten (Williams 2012, Table  1 therein). Thus, unless fluxed spectra are available at each photometric point to assess the contribution of the H$\alpha$ and OI lines to the broad band fluxes, it becomes difficult to deconstruct and interpret the R$_{c}$-I$_{c}$ color evolution. One may cautiously conjecture that the increasing trend of the R$_{c}$-I$_{c}$ color from JD 2457350 onwards is due to a dominating contribution from the Balmer H$\alpha$ line whose strength can decline slower than the  continuum itself (Munari et al. 2015).

\subsection{Optical Spectroscopy}
\label{subsec-OptSpec}
\begin{figure}
\centering
\includegraphics[bb=3 43 426 578, width= 3.5in,clip]{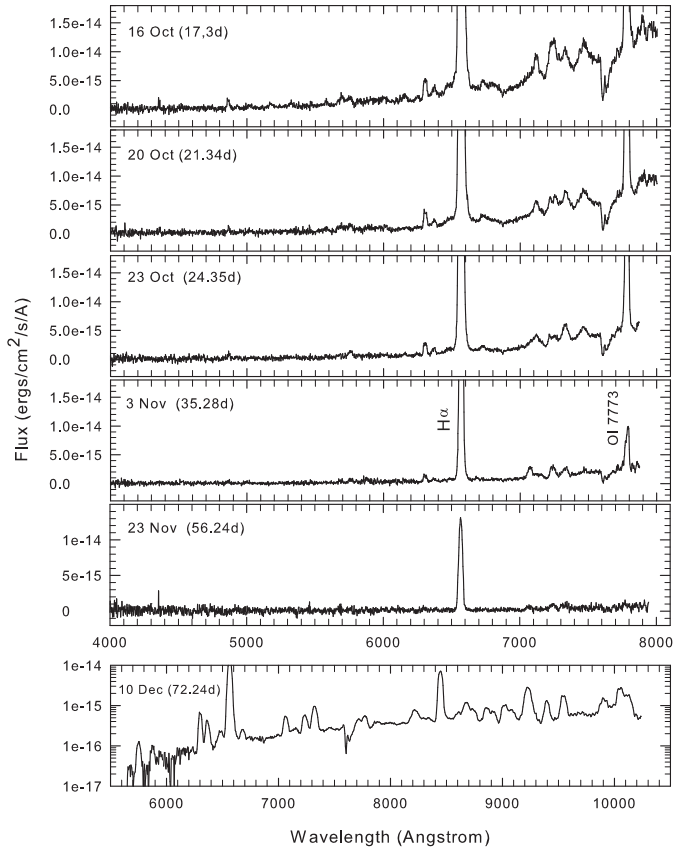}
\caption[]{Optical spectra of V1831 Aql at selected epochs are shown. Observation dates are indicated, as well as the days after maximum which are shown in brackets.}
\label{fig-OptSpec}
\end{figure}

\begin{figure}
\centering
\includegraphics[bb=33 144 547 671, width= 3.5in,clip]{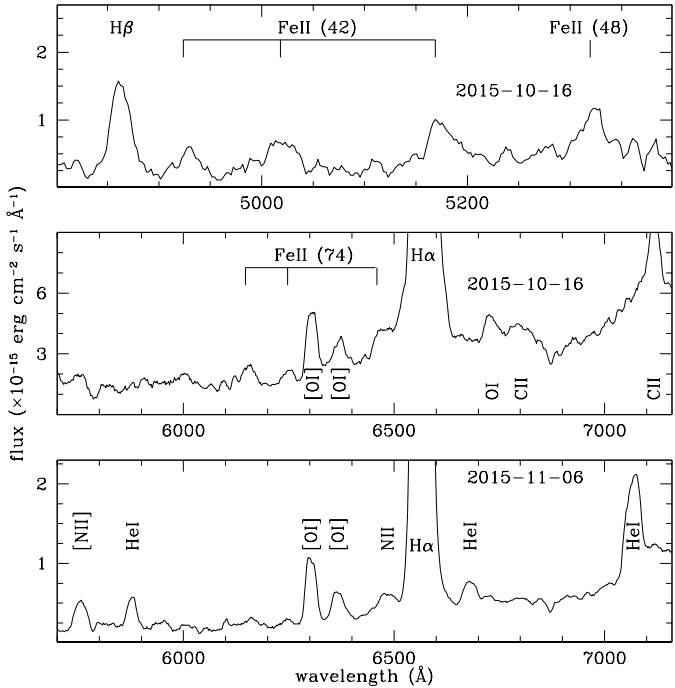}
\caption[]{Magnified view of selected sections of the optical spectra are shown to establish the Fe II class of the nova and also show a few other spectral  characteristics discussed in the text.}
\label{fig-Opt-expanded}
\end{figure}

The low resolution optical spectra are presented in Figure~\ref{fig-OptSpec}.  The rise of the continuum in  Figure~\ref{fig-OptSpec} toward the red shows the large reddening towards the object which also manifests itself in the blue end of the spectrum where the signal in the emission features is  considerably suppressed. An expanded  view of selected  regions of the spectra of 2015 Oct 16 and 2015 Nov 06  (these have  the highest S/N), after a five point boxcar smoothing to suppress noise and enhance visibility of weak lines is shown in Figure~\ref{fig-Opt-expanded}. In the  top panel of Figure~\ref{fig-Opt-expanded}, lines of the Fe II (42) and (48) multiplets are clearly seen thereby establishing the nova to be of the Fe II class. The middle panel, for the same date, looks at the region around H$\alpha$. Here again lines of Fe II and CII are seen, very typical of the  Fe II class of novae, and no helium or nitrogen lines typical of William's (1992) He/N type are seen. Under sufficient magnification, many weak lines are  better seen indicating that the spectra of October are closely similar to that of a typical Fe II nova around maximum light and early decline (for e.g, there is a good similarity with the spectra of the Fe II nova V977 Sco  shown in Fig. 5 of Williams et al. (1991)). The bottom panel of Figure~\ref{fig-Opt-expanded} plots the same region as the middle one, to highlight the disappearance of Fe II and CII lines, and show the  emergence of HeI and NII lines, indicative of an evolution toward higher ionization conditions, as typical of Fe II novae. The appearance of the auroral [NII] 5755$\AA$ line always precedes the emergence of the nebular [NII] 6548, 6584$\AA$ doublet. Whether this doublet is already present in emission we cannot say from this spectrum given the large width of H$\alpha$, whose wings overlap and extend beyond the position of the doublet. The [OII] 7319$\AA$, 7330$\AA$ auroral lines, which generally accompany the [NII] 5755$\AA$ line can be also seen in the spectrum of 2015 Dec 10 (Figure~\ref{fig-OptSpec}, bottom panel). The strongest non-hydrogen line, through  the evolution shown here, has been an oxygen line. It is either the OI 7774$\AA$ line or the Lyman $\beta$ fluoresced 8446$\AA$ OI line. Thus, following the spectral phase classification scheme proposed by Williams et al (1991), an appropriate  spectral phase  that could be assigned to the nova, over the evolution shown here, would be $P_{o}$. It is difficult to comment on the addtional spectral features considering the very low signal to noise ratio (SNR) of the spectra of this faint nova.
\par
\begin{figure}
\centering
\includegraphics[bb= 2 0 399 297, width= 3.0in,clip]{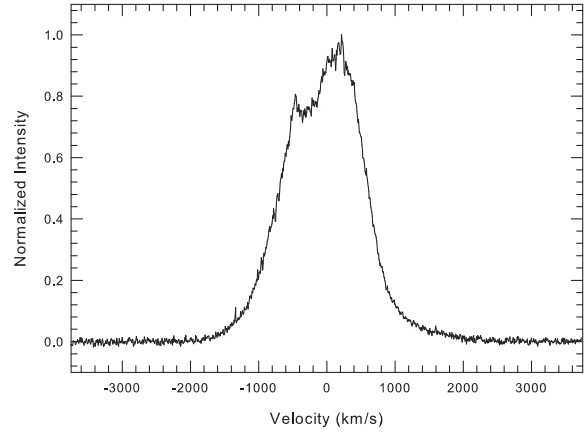}
\caption[]{The velocity profile of the H$\alpha$ 6563\AA~~ line on 2015 October 23.85  obtained at a resolution of 20,000. The
FWHM of the profile is  1300 km/s.}
\label{fig-HAlpha}
\end{figure}

A high resolution (R = 20000) echelle spectrum of the H$\alpha$ line is presented in Figure~\ref{fig-HAlpha}. The half width at zero intensity (HWZI) of the line is $\sim$ 2000 km/s which agrees with its Fe II type. As per the classification scheme of  Williams (1992),  the observed HWZIs of the Fe II novae are generally  $<$ 2500 km/s while for the He/N class it is $>$ 2500 km/s. Correlations are known to exist  between the outflow velocities of the ejecta with the speed class. Mclaughlin (1960) gives for the velocity $v$ of the principal ejecta the relation log($v$) = 3.57 - 0.5$\times$log $t_{2}$ which for $t_{2}$ = 16.4 $\pm$ 0.5d yields an expansion velocity of 917 $\pm$ 13 km/s. This is smaller than the observed FWHM of 1300 km/s. On the other hand, studies of novae in M31 by Shafter et al. (2011) yields a similar relation viz log $t_{2}$ (d) = (6.84$\pm$ 0.10) - (1.68 $\pm$ 0.02) log ($FWHM_{Halpha)}$ which leads to  an FWHM value of 2230 $\pm$ 270 km/s for the given $t_{2}$. This is on the higher side compared to what is observed. The difference between the  observed and  outflow velocities inferred from the two empirical laws may be reconciled, in part at least, by a combination of  factors. As Shafter et al (2011) point out, a major factor in the discrepancy between their results and those of McLaughlin (1960) is that in the latter relation the expansion velocities are derived from the absorption-line minima of the P Cyg profiles measured near maximum light. Such velocities, Shafter et al (2011) point out,  are only 20\% to 50\% of those inferred from the emission-line FWHM. On the other hand, Shafter et al. (2011) attribute the scatter in their data, particularly for the slower novae, as probably arising in part from the time dependence of the derived velocities which were obtained from spectra taken at varying times after outburst.
\par
The low resolution spectra are useful to determine the neutral OI gas mass and temperature. Williams (1994) has shown that the observed flux ratio I(6300)/I(6364) of the forbidden lines of neutral OI is invariably smaller than the value of 3:1 expected from their transition probabilities if these lines were optically thin. This suggests that the OI lines are optically thick which is not expected based on the relatively small abundance of neutral oxygen  in the ejecta and the relevant atomic parameters for the line transitions. The neutral OI, it is therefore suggested, must reside mostly in dense globules where it is protected from the harsh ionizing radiation from the central white dwarf. Following Williams (1994), the optical depth $\tau$ in the [OI] 6300\AA line, the temperature T$_{e}$ of the neutral gas and the mass M$_{O I}$ of the neutral OI can be respectively found from applying the following three equations in succession:

\begin{equation}
\frac{I_{6300}}{I_{6364}} =  \frac{1-e^{-\tau}}{1-e^{-\tau/3}}
\end{equation}

\begin{equation}
T_{e} =  \frac{11,200} {log(43{\times}R{\times}\frac{\tau}{1-e^{-\tau}})}K
\end{equation}

\begin{equation}
M_{O I} = 152d^{2} e^{22850/T_{e}} \times 10^{1.05E(B-V)}\frac{\tau}{1 - e^{-\tau}} I(6300) M_\odot
\end{equation}

where $R$ is the observed I(6300)/I(5577) flux ratio and $d$ is the distance in kilo-parsecs (kpc). Because the measurement of the strength of the 5577\AA ~line is prone to error due to its weakness, we have limited our analysis to the spectrum of 2015 October 16 in which this line is best seen. From the spectrum, after correcting  for reddening using $E(B-V)$ = 2.92, we measure I(6300)/I(6364)= 1.86 $\pm$ 0.10 and I(6300)/I(5577) = 1.62 $\pm$ 0.43 (reddening corrections were made using $A$(5577) = 0.9831$A_{v}$, $A$(6300) = 0.8585$A_{v}$ and $A$(6364) = 0.8485$A_{v}$ from Cardelli et al. (1989)). This yields $\tau = 1.77 \pm 0.24$, $T_{e} = 5156 \pm 170K$ and $M_{OI}$ = $(8.8 \pm 2.2)$ $\times$ $10^{-5}$ $M_{\odot}$. These values agree well with the list of tabulated values for a large number of novae given by Williams (1994).

\subsection{Results from near infrared observations}

The log of the spectroscopic observations is shown in Table~\ref{table-NIRSpec}. The evolution of this nova can be broadly divided, for ease of discussion,  into an early phase comprising of the spectra between 8 October to 19 October 2015. Spectra subsequent to this belong to  a phase which is dominated by emission from  dust.

\subsubsection{Early phase spectra}

\begin{figure*}
\centering
\includegraphics[bb= 32 240 519 654, width = 5.5in,clip]{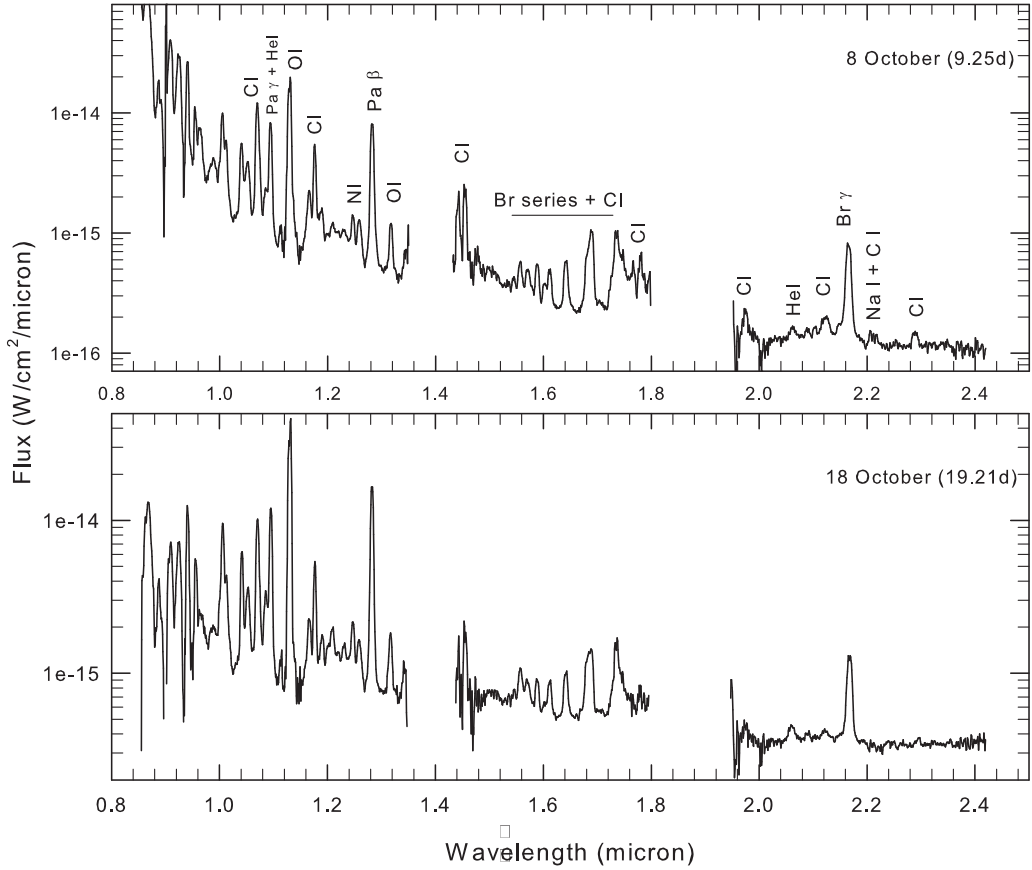}
\caption[]{Two representative spectra from the early stage with the prominent lines  marked.}
\label{fig-NIRSpec}
\end{figure*}

Clear evidence for dust nucleation in the nova was seen by end of October 2015. Two representative spectra  are shown in Figure~\ref{fig-NIRSpec} with the principal lines identified. A more detailed identification table can be found in Das et al (2008) and in Banerjee $\&$  Ashok (2012) where templates of the NIR spectra of several Fe II and He/N novae are given. The spectra are typical of the Fe II class, where apart from  prominent lines of HI, OI and HeI, several strong lines of neutral Carbon are seen. These lines, which distinguish the Fe II from the He/N class of nova in the infrared, made the Fe II classification a simple exercise in the present case, in an instance where an optical classification was relatively difficult.  The Pa$\beta$ and Br$\gamma$ HI lines have FWHMs of 1300 to 1350 km/s, after  deconvolving for instrument broadening, which matches well the FWHM of 1300 km/s obtained from the high-resolution H$\alpha$ echelle profile. Among other notable features of the spectra are the presence neutral NaI lines (e.g 2.2056 and 2.2084 $\mu$m) whose presence had inevitably preceded dust formation (see Das et al. 2008). In this instance too, dust did indeed form in the nova.
\par
No first overtone CO emission bands at 2.29$\mu$m and beyond were seen in any of the spectra. From the known first overtone CO detections in novae, compiled in Banerjee et al. (2016), CO emission in novae is generally seen during the first 1 to 3 weeks after outburst commencement. CO emission is a transient and short lived phase that generally does not last more than 2 weeks (Banerjee et al. (2016) and references therein) consistent with theoretical models for CO formation in nova winds (Pontefract $\&$ Rawlings, 2004). The transient nature of the emission (Das et al. 2009) makes it easy to miss and it is possible, though we deem it unlikely, that CO was formed and destroyed in this nova during the gaps in our observations (for e.g. during the first 17 days after discovery when we have only 2 epochs of spectroscopic data separated by $\sim$ 8-9 days). All  novae known to have shown  CO emission have invariably proceeded to form dust but the converse is not true. Dust has formed in novae where there has been no CO formation. An example is the bright nova V1280 Sco, which formed copious dust (Das et al. 2008; Chesneau et al. 2008; Chesneau et al. 2012), but where no CO emission was detected. NIR spectroscopic observations of V1280 Sco up till the dust forming stage, were taken at high and sufficiently regular cadence to detect CO had it formed notwithstanding its transient nature (Das et al. 2008).

\subsubsection{Electron density, emission measure and mass estimate}
\label{subsubsec-ElecDensity}

We measured the dereddened Brackett line strengths for the  spectrum of 8 October 2015 and performed a Case B analysis. This spectrum was chosen because dust formation is unambiguously absent at this stage; the measured H line strengths are thus not complicated by any effects/contribution from  dust emission. Since the analysis presented  here, is identical to that done for novae Nova Sco 2015 and Nova Cep 2014 (Srivastava et al. 2015), the details of the present analysis are not elaborated upon; rather the main findings are  summarized. The Br$\gamma$ 2.1656 $\mu$m line is found to be optically thick as its strength with respect to other Br lines ( Br 12 to 16) deviates from Case B predictions for any reasonable combination of temperature $T$ and electron density $n_e$. For e.g. the Br$\gamma$/Br14 ratio is found = 4.15 $\pm$ 0.22, whereas the expected Case B value is 7.63 $\pm$ 2.50 when averaged over three different temperatures of 5000K, 10000K and 20000K and considering an extended density range between $10^{8}$ to $10^{14}$ cm$^{-3}$ for each temperature. Even when the temperatures are considered separately, the observed Br$\gamma$/Br14 ratio is significantly lower than the expected Case B values. The high density range is considered because high values of the density  are expected in the ejecta just 9.25d after the eruption. Given that nova shell masses  are typically estimated to be in the range $10^{-4}$ to $10^{-6}$ $M_{\odot}$ (e.g. Gehrz 1988;  Della Valle et al. 2002), it is  straightforward to show that  during the early stages after outburst, $n_{e}$ must be high and lie in the middle of the extended range considered. We use the constraint that the optical depth at the line center of the Br$\gamma$ line $\tau_{(Br\gamma)}$ is greater than  1,  where following Hummer \& Storey (1987) and Storey \& Hummer (1995),

\begin{equation}
\tau_{(Br\gamma)} = n_e n_i L \Omega_{(Br\gamma)} > 1
\end{equation}

where $n_e$, $n_i$, $L$ and $\Omega_{(Br\gamma)}$ are the electron number density, ion number density, path length and Br$\gamma$ line opacity respectively. $L$ is taken as  the kinematical distance $v \times t$ traveled by the ejecta where $v$ is the velocity of ejecta (assumed = 650 km/s) and $t$ =9.25d is the  time after discovery. As $\tau_{(Br\gamma)}$ can be $> 1$, the lower limit on the electron density $n_e$ is found to be  in the range $\sim$ $5\times 10^9$ cm$^{-3}$ to $1.2 \times 10^{10}$ cm$^{-3}$ (assuming $n_e = n_i$) and the emission measure $\sim$  $4.5\times 10^{33}$ cm$^{-5}$. The gas mass of the ejecta may be estimated by $M$ = $\epsilon$$V$$n_e$$m_H$ where $V$ is the volume ( = 4/3$\pi$$L^{3}$), $\epsilon$ is the volume filling factor  and $m_H$ is the proton mass. Using the lower limits on $n_e$ as estimated above, the lower limit on the mass $M$ varies between $(2.4 - 5.7)\times 10^{-6}$  M$_{\odot}$ assuming $\epsilon$ =1.

\begin{table}
\centering
\caption{Evolution of the (HeI 1.083 $\mu$m)/(Pa $\gamma$ 1.094 $\mu$m) and (HeI 2.058$\mu$m)/(Pa $\gamma$ 1.094$\mu$m) line strength ratio}
\begin{tabular}{lcccccc}
\hline
Date of    & Time  &(HeI 1.083)/(Pa$\gamma$ 1.094)  &  (HeI 2.058)/(Pa$\gamma$ 1.094)             \\
Observation& (days)& ratio \\
(dd/mm/yy) &       &       \\
\hline
\hline
08/10/16 &  9.25 & 0.31	& 0.01 \\
16/10/16 & 17.22 & 0.28	& 0.04 \\
17/10/16 & 18.18 & 0.25	& 0.02 \\
18/10/16 & 19.21 & 0.32	& 0.02 \\
19/10/16 & 20.22 & 0.31	& 0.01 \\
27/10/16 & 28.17 & 2.40	& 0.16 \\
04/11/16 & 35.18 & 4.78	& 0.19 \\
05/11/16 & 36.18 & 5.38	& 0.06 \\
19/11/16 & 51.13 & 7.19	& 0.28 \\
26/11/16 & 58.11 & 5.55	& 0.07 \\
02/12/16 & 64.07 & 5.45	& 0.20 \\

\hline
\hline
\end{tabular}
\label{table-HeIPaBeta}

\begin{list}{}{}
\item
\end{list}
\end{table}

\subsubsection{The HeI 1.0830 $\mu$m line strength evolution with respect to the Pa$\gamma$ 1.094 $\mu$m line}
\begin{figure}
\centering
\includegraphics[angle=0,width=0.45\textwidth]{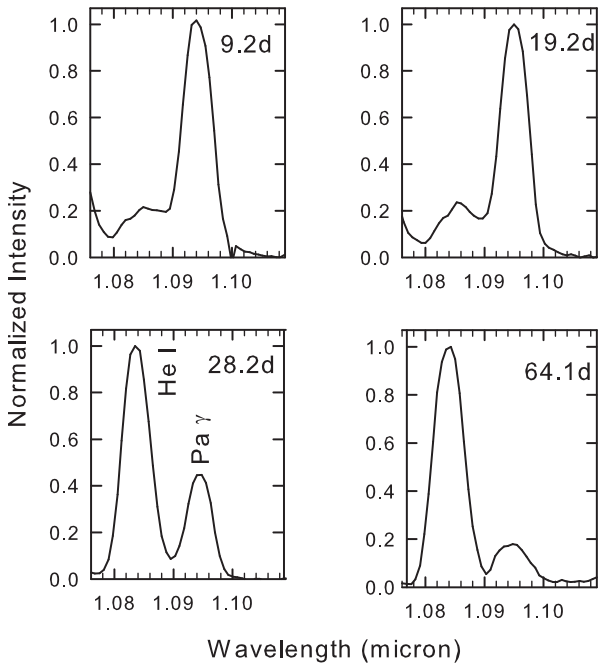}
\caption[]{The evolution in the strengths of the HeI 1.083 $\mu$m and Pa $\gamma$ 1.094 $\mu$m line with time after discovery indicated in days.}
\label{fig-HeIPaBeta}
\end{figure}

One of the striking features of the NIR spectra is the rapid evolution of the (HeI 1.083)/(Pa$\gamma$ 1.094) ratio (denoted by $R$ henceforth) during the early stages. This evolution is shown in Figure~\ref{fig-HeIPaBeta} and Table~\ref{table-HeIPaBeta}. The  HeI 1.0830 $\mu$m line strength strengthens rapidly and the sharp increase in the value of $R$, specifically between 20d to 28d when it  changes from 0.31 to 2.4, is quite remarkable. The early evolution of a large number of novae of the Fe II class have been followed from Mt. Abu (Banerjee $\&$ Ashok, 2012 and references within) but we have not encountered a case where the HeI 1.0830 $\mu$m line brightens so rapidly and persists at a high intensity at such an early stage after outburst. The only exception was nova V574 Pup but this can be explained by the consideration that it was a hybrid nova (Naik et al 2010) of the Fe IIb type (Williams 1992) where strong He emission is expected after the transition from the Fe II to the He/N class has been made (strong He lines are expected by definition in He/N class novae). V754 Pup made such a transition approximately a month after its outburst. Having said the above, it should also be mentioned, that HeI 1.0830 $\mu$m can become the strongest line in NIR spectrum, by a very large margin with respect to other lines, but {\it only} during the later nebular and coronal stages. Examples of such novae, with overwhelmingly strong HeI 1.0830 in emission, are PW Vul (Williams et al. 1996), V574 Pup (Naik et al. 2010), Nova Mon 2012 (Banerjee et al. 2012) and V5668 Sgr (Banerjee et al. 2016; Gehrz et al. 2016).
\par
In PW Vul, the ratio $R$ increased by a factor of about 70  from $\sim$ 0.4  on day 78 to $\sim$ 27  on day 272, at a phase when the coronal lines were conspicuous. Williams et al. (1996) discussed the possible cause for this after considering the effects of radiative and collisional processes on the observed line ratio $R$. They find that for between 1250K and 20000K, the ratio $R$ is approximately proportional to the temperature (as a consequence of collisional processes playing a dominant role and with the relevant collision strengths having a significant temperature dependence; see their equation 6). They thus suggested that in case of nova PW Vul, the large 70 fold increase in $R$, seen between 78d to 272d after outburst, was partly due to the rising temperature in the region where the He/H emission originated. In essence, at least part of the observed behavior of $R$ is a consequence of the increasing excitation of the nova. This seems to be a viable reason for the later nebular/coronal stages when  excitation conditions increase in the ejecta. However, for the behavior observed here - very early after outburst -  we offer an alternative explanation. We believe that the HeI 1.083 $\mu$m line was optically thick on day 9.25 (when $R$ = 0.31)  and was hence seen at greatly reduced intensity. The electron density at around this time was at least $10^{10}$ cm$^{-3}$  or greater as the preceding Case B analysis showed. The emission line spectrum of neutral Helium has been calculated by Almog $\&$ Netzer (1989) up to high densities of $10^{14}$ cm$^{-3}$. They show that at densities exceeding $10^{10}$ cm$^{-3}$ the HeI 1.0830 $\mu$m line emissivity and line strength dramatically drop as the optical depth increases (see Figures 1 and 3 of Almog $\&$ Netzer (1989)). Thus by day 28.17, if the ejecta dilutes geometrically with $n_e$ $\propto$ to $t^{-2}$ or ballistically with $n_e$ $\propto$ $t^{-3}$, $n_e$ may be expected  to reduce by a factor of $\sim$ 10 to 27 and it is quite possible that the ejecta changes from optically thick to optically thin (or partially thin) conditions between day 9.25 to day 28.17. In which case the line strength should go up considerably as is indeed seen on day 28.17 when $R$ = 2.4.
\par
The second strongest HeI line in the NIR region viz the 2.0581 $\mu$m line also follows a similar behavior as the 1.083 $\mu$m line. Unfortunately the 2.0581 $\mu$m line lies in a window of poor atmospheric transmission due to strong $CO_2$ absorption, it is considerably weaker than the 1.083 $\mu$m line  and further it gets veiled as dust formation begins to contribute to the IR excess in that region. As a result the signal to noise ratio (SNR) in this line is low in most of our spectrum which hampers drawing strong conclusions about its evolution. But even then, as seen from the last column of Table~\ref{table-HeIPaBeta}, there is a strong and sudden increase in its strength between day 20 to day 28 by more than a factor of 10 with the increased strength tending to persist subsequently. This behavior is quite similar to that of the 1.083 $\mu$m  line. However, unlike the 1.0830 $\mu$m line, Almog $\&$ Netzer (1989) have not analysed the behavior of the 2.0581 $\mu$m line  at high densities so we cannot be sure that we are witnessing optical depth effects here too. Generally, the strength of the 2.0581$ \mu$m line has been computed mostly at low densities. Benjamin et al. (1999) present line strengths for typical nebular conditions for densities up to $n_{e} = 10^{6}$ cm$^{-3}$. We believe there are certain complexities in calculating its line strength at higher densities (R. Porter, private communication); it is not included in the computed strengths of HeI lines by Porter et al. (2005).
\par
The question may arise why similar behavior of the HeI lines is not seen in other novae. A simple explanation eludes us. But there are atleast two other novae worth discussing viz. recurrent nova T Pyx and V5558 Sgr where the HeI line behavior, if not identical to this nova, was rather unusual. T Pyx, during its 2011 outburst, showed  a significant transition of the nova from the He/N to the Fe II class, from both NIR and optical spectra, within a few days after the outburst (Joshi et al. 2014; Ederoclite et al. 2013). That is, it belonged to the hybrid class of novae (Williams 1992) but with the transition occurring in the reverse direction (He/N to Fe II rather than Fe II to He/N). Prominent HeI lines were seen soon after the onset of the eruption which then faded rapidly in strength within the next few days. A similar behavior was also seen in the optical in V5558 Sgr (Tanaka et al. 2011). The reason for this phenomenon is not well understood.


\begin{figure*}
\centering
\includegraphics[bb= 38 420 545 815, width= 5.5in,clip]{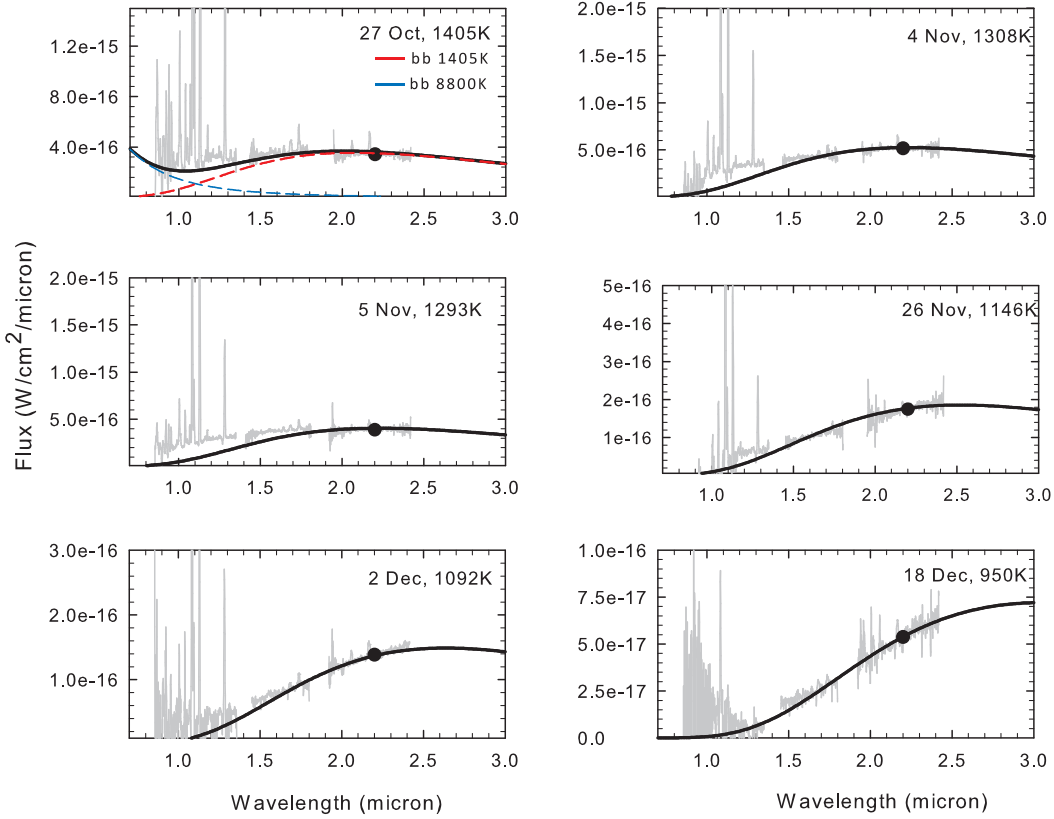}
\caption[]{Selected JHK spectra (grey) during the dust formation phase along with their blackbody fits (black) and temperatures indicated. The windows of poor atmospheric transmission between 1.4 to 1.5 microns and 1.8 to 1.95 microns are blanked out. The blackbody fits show a deficit in the $J$ band for the earlier epochs which can be accounted for by using an additional contribution from a hotter blackbody at around $\sim$ 8000-9000K from the central remnant. For example, we show  the co-added sum (black line) of a two-component gaussian fit for the data of 27 October in the  top left panel where the hot component (blue dashed line) has insignificant contribution in the $K$ band region where the dust blackbody (red dashed line) dominates overwhelmingly.  }
\label{fig-BBFit}
\end{figure*}

\begin{figure}
\centering
\includegraphics[bb= 2 42 398 251, width= 3.0in,clip]{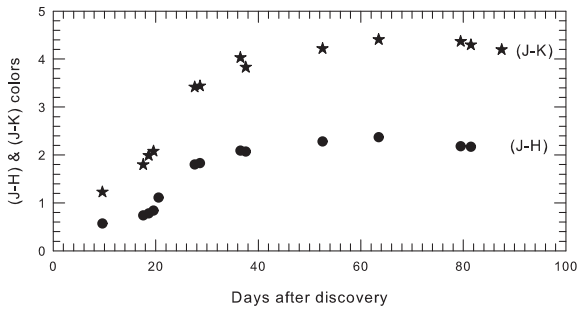}
\caption[]{Evolution of the near-infrared $(J-H)$ and $(H-K)$ colors with time during the dust formation phase }
\label{fig-JHKColor}
\end{figure}

\begin{figure}
\centering
\includegraphics[bb= 115 307 523 724, width= 3.25in,clip]{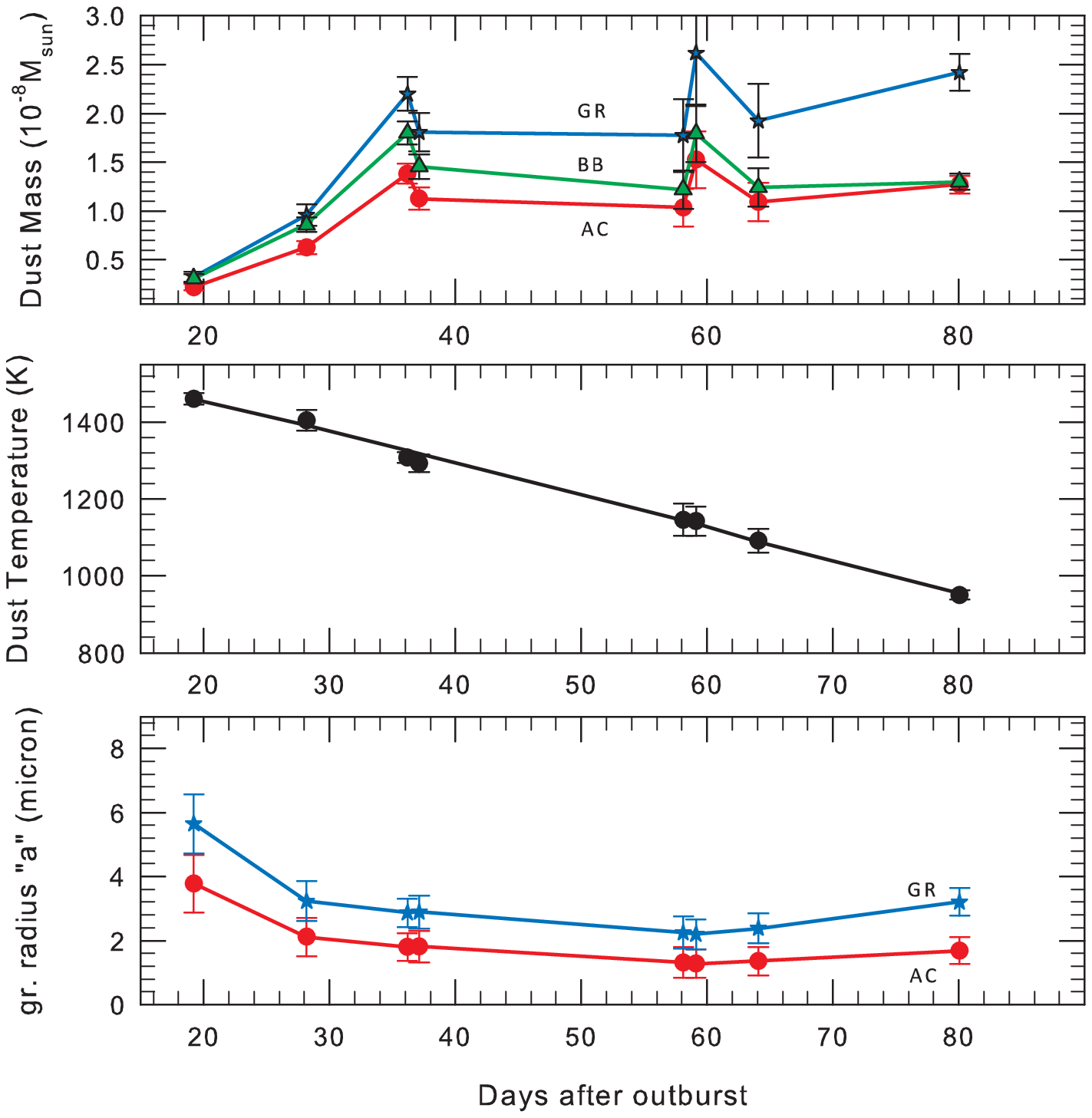}
\caption[]{Evolution of the dust mass, temperature and grain radius with time. A detailed discussion of the evolution of these parameters
has  been made in the text. AC = amorphous carbon, GR = graphite and BB= blackbody.}
\label{fig-DustPara}
\end{figure}


\subsubsection{Dust formation and evolution}
The evolution of the IR spectra is shown in Figure~\ref{fig-BBFit} where  an IR excess is clearly seen at longer wavelengths  that builds up with time. We attribute this  excess to dust emission and model it through blackbody fits. NIR photometry also shows independent supporting evidence for dust formation from the evolution and gradual increase in the $(J-H)$ and $(J-K)$ colors during this time  (Figure~\ref{fig-JHKColor}). From Figure~\ref{fig-JHKColor} it appears that grain nucleation may have begun even as early as day 17-19 after discovery because the $(J-H)$ and $(J-K)$ colors had already begun to increase at this stage. We show shortly that spectroscopy confirms this. Thus this appears to be  a fairly early detection of dust formation in a nova; the earliest known instance occurring in Nova Her 1991 (V838 Her; Harrison $\&$ Stringfellow 1994) at +8d after outburst. Williams et al (2013) have analysed the correlation between the dust formation timescale and the speed class from observational data of CNe and show that dust typically takes longer to form in slower novae. For a $t_{2}$ of $\sim$ 16.5d as estimated here, a value of approximately 25-27 days for the onset of dust condensation is suggested from their correlation plot. This is in  reasonable agreement with the dust condensation time of +19.2d that we observe considering that there is statistical scatter in the  Willliams et al. (2013) correlation plot. What is perhaps more important is  the qualitative agreement  that, among dust forming novae, the fast novae proceed to form dust quickly.
\par
The blackbody fits for the spectra starting from 27 October (+28d) are discussed first. The earliest epoch at +19.2d when dust formation had just commenced is discussed and analysed separately later. The blackbody fits of Figure~\ref{fig-BBFit} give good agreement to the observed data at longer wavelengths (viz. $K$ band) but there is a deficit in the $J$ band. We have checked that this deficit  can be accounted for by using an additional contribution from a hotter blackbody (at around $\sim$ 8000-9000K) from the central remnant.For example, we show  a two-component gaussian fit for the data of 27 October (+28d) in the  top left panel of Figure~\ref{fig-BBFit}. As is clearly established, the hot component has insignificant contribution in the $K$ band region where blackbody representing the dust emission dominates overwhelmingly. As a consequence of this,  it is justified in neglecting the contribution from the hot component and we deem it sufficient to give  primary importance to the $K$ band data while arriving at  best fits for the  dust emission to estimate the dust temperature. The best fits were obtained using a standard least squares minimization Marquardt-Levenberg algorithm;  the formal errors are small because of the large number of points in the $K$ band that are being fit. A limitation of this work is the lack of data at longer mid-IR wavelengths that would have given a more extended SED and more reliable estimates of the dust temperature to be made.
\par
The dust temperature and the $(\lambda{f_{\lambda})}_{max}$ values are listed in Table~\ref{table-DustPara}. Using these we estimate how optically thick the dust was by comparing the dust and outburst luminosities. Given that the integrated flux under a blackbody curve is given by 1.35$\times$$(\lambda$${f_{\lambda})}_{max}$ (Gehrz $\&$ Ney 1992) we have the maximum dust luminosity $L_{IR}$ =  1.2$\times 10^4 L_{\odot}$ on day 36.37 using $(\lambda {f_{\lambda})}_{max}$  from Table~\ref{table-DustPara}. The outburst luminosity is estimated from $M_{bol}$ = 4.8 + 2.5log (L/$L_{\odot}$) where the bolometric correction applied to $M_V$ is assumed to lie between -0.4 and 0.00 corresponding from A to F spectral types respectively (novae at maximum generally have a spectral type between A to F). Using $M_V$ = -8.29 as deduced previously, the outburst luminosity $L_O$  is then (2.11 $\pm$ 0.38)$\times$ $10^5$ $L_{\odot}$ which is approximately 6 to 7 times the Eddington luminosity for a 1 solar mass white dwarf. We adopt a mean value of 2.11$\times$ $10^5$ $L_{\odot}$ for the outburst luminosity in subsequent calculations. In comparison,  the dust luminosity $L_{IR}$ is significantly less with $L_{IR}$/$L_O$ being $\sim$ 0.06 indicating that very little of the central engine's  optical/UV output radiation is intercepted, absorbed and reprocessed into IR emission. This would imply that the dust could either  be clumpy and hence does not cover all  angles of the sky towards the nova as seen by the observer. Alternatively the dust shell may be homogenous but having insufficient material to be optically thick i.e it does not behave as an ideal calorimeter that traps all radiation. The few points available in the $V$ band light curve between 10 to 20 November 2015 (at a time when dust was present) have large error bars but still do not show any hint of a strong dip in optical brightness as expected for an optically thick dust forming event e.g. similar to the deep dips in the dust forming events in V705 Cas (Evans et al. 1996), V5668 Sgr (Banerjee et al. 2016) or  a few other D-type (dust type) novae catalogued in Strope et al.(2010).
\par
The behavior here is similar to two other optically thin dust forming novae  Nova Cyg 2006 (Munari et al. 2008) and V2615 Oph (Das et al 2009). In the former nova  it was  found (Munari et al. 2008) that the optical depth of the dust reached 0.2 with no measurable effects on the lightcurve at optical wavelengths but displaying however a large increase in the  $JHK$ magnitudes. The mass of the ejecta was estimated as 5$\times$10$^{-4}$ M$_\odot$ from [OI] lines and 3$\times$10$^{-4}$ M$_\odot$ from CLOUDY modeling, while the IR emission from dust suggested that $M_{gas}$/$M_{dust}$= 3$\times$10$^{5}$.

\subsubsection{Dust mass and grain size}

To estimate the dust mass  it is assumed  that  the grains  are spherical with a mean radius $a$ and that the dust is composed of carbonaceous material.  The justification for the carbon composition is discussed  shortly. Based on the mean Planck absorption efficiencies $<$$Q_{abs}$$>$  for carbon grains provided by Blanco et al. (1983), Evans et al. (2016) have shown that $<$$Q_{abs}$$>$  can be well approximated by a form $<$$Q_{abs}$$>$ = $A a^{\gamma}$ $T^{\beta}$. They find the Planck mean absorption efficiencies from Blanco et al. (1983) are very well fit by the relations $<$$Q_{abs}$$>$/$a$ = $53.160$ $T^{{0.754}}$ and $<$$Q_{abs}$$>$/$a$= $0.653$ $T^{{1.315}}$ for amorphous carbon (AC) and graphitic (GR) grains respectively with $a$ in cm and T in Kelvin. The fit for the AC grains is excellent over the temperature range 400 - 1700 K while for graphitic carbon the fit is reasonably satisfactory over the range 700 - 1500K (i.e. over the range of  dust temperatures  seen here in V1831 Aql; refer Table~\ref{table-DustPara}). For blackbody grains $A$ = 1 and $\beta$ = $\gamma$ = 0. Using these relations, it can be shown (Evans et al. 2016) that the dust masses for AC, GR and blackbody  grains are given by:

For optically thin  amorphous carbon,
\begin{equation}
\frac{M_{dust-AC}}{M_{\odot}} \simeq 6.91 \times 10^{17} \frac{(\lambda f_{\lambda})_{max}}{T^{4.754}_{dust}}
\end{equation}

and for optically thin graphitic carbon,
\begin{equation}
\frac{M_{dust-GR}}{M_{\odot}} \simeq 6.16 \times 10^{19} \frac{(\lambda f_{\lambda})_{max}}{T^{5.315}_{dust}}
\end{equation}

and if the grains behave as perfect blackbodies

\begin{equation}
\frac{M_{dust-bb}}{M_{\odot}} = 4.02 \times 10^{15} \frac{ a(\lambda f_{\lambda})_{max}}{T^{4}_{dust}}
\end{equation}

where a value of $D$ = 6.1 kpc has been used for the distance,  $\rho$ = 2.25 gm cm$^{-3}$ has been taken for  the density of  the carbon grains, grain  radius $a$ = 1 $\mu$m is assumed for the black body dust mass relation and ($\lambda$ $f_{\lambda}$)max is in units of W/m$^{2}$. Estimated dust masses are tabulated in Table~\ref{table-DustPara} using the T$_{dust}$ values listed therein. It must be noted that the entry in Table~\ref{table-DustPara} on 19.2d for the temperature is not estimated from a blackbody fit but done slightly differently; this procedure is now described. It is first established that a mild but definite IR excess was present on 19.2d which was not present  in the first spectrum of +9.2d. The top panel of Figure 12 shows the spectra on days +9.2 and +19.2 with the spectrum of day 19.2  scaled upwards  to match the spectrum of 9.2d in the $J$ band. As a consequence of this exercise a clear IR excess is seen to have newly developed in the $H$ and $K$ bands. The same conclusion is also reached by plotting the dereddened R$_{c}$,I$_{c}$,J,H,K$_{s}$  fluxes from this paper on days 9.2 and 19.2 in the bottom panel. The $J$ band fluxes of both days have been matched after scaling. A newly developed IR excess is seen in the $H$ and $K$ bands while the fluxes in all other bands match well indicating the nova has faded reasonably similarly in these bands. It is clear that dust has begun to contribute to the SED on day 19.2. It is not meaningfully possible to fit a blackbody to this dust excess to estimate the dust temperature without  causing substantial errors because the excess is mild, and the contribution of the hot central  remnant (as discussed earlier) to the $K$ band is uncertain but could be significant. We thus estimate the dust temperature  on +19.2d to be  1461 $\pm$ 15K  from  extrapolation of the linear  best fit to the dust evolution curve shown in the middle panel of Figure 11. An upper limit on the dust flux  can be set  by attributing that the entire $K$ band flux on day +19.2 is due to dust emission. In which case,  a blackbody curve at 1461K  drawn to match the $K$ band flux level of 19.2d yields  a value of ($\lambda$$f_{\lambda}$)max =  3.49$\times$$10^{-12}$ W/m$^{2}$ which is listed in Table~\ref{table-DustPara}. This value of ($\lambda$$f_{\lambda}$)max, as well as the mass estimate derived from it using equations 5 to 7, are upper limits and are indicated so in Table~\ref{table-DustPara}.
\par
Regarding the choice of carbon for the dust composition, Lodders and Fegley (1995) and Lodders and Fegley (1999) have shown that  the condensation temperature of different minerals is very  much dependent on the C/O ratio, the gas pressure and elemental abundances. However, most estimates show  that the condensation temperature for Carbon dust (both amorphous and graphite) is high and lies in a range between 1600 - 1800K in most astronomical environments such as the ISM (Blanco et al. 1983 ); in Type II supernovae (Todini and Ferrara, 2001) or  C rich stellar environments of M and N stars (Lodders and Fegley, 1995, 1999). Condensation temperatures for graphite may even exceed 1800K under certain conditions (Lodders and Fegley, 1995). The above mentioned studies  similarly show that most silicates condense at lower temperatures between 1100 - 1400K (Todini and Ferrara, 2001; Speck et al. 2001;, Lodders and Fegley, 1995, 1999). The dust temperature at the first epoch of  1461K is too high to favor a silicate composition but consistent with  a carbonaceous constitution. A carbon composition is also consistent with the presence of numerous strong lines of neutral CI as seen in the NIR spectra in Figure~\ref{fig-NIRSpec}. The treatment of a Si rich composition is beyond the scope of this work.
\par
Similar or higher temperatures for the dust, as observed here, have been estimated in several novae especially when the  dust emission  has been detected early due to  monitoring at high cadence. For e.g. a temperature of 1615K was detected in the Fe II nova V476 Scuti (Das et al. 2013), which might have eluded detection had the observation been untimely since the dust rapidly cooled to 1040K within 24 days. The first dust detection in nova V339 Del was at 1637K (Evans et al. 2016),  temperatures in the range 1300-1700K were estimated in V1280 Sco at different stages of the dust evolution (Chesneau et al. 2008) while the dust in the classical nova V2326 Cyg condensed at a temperature 1410 $\pm$ 15 K (Lynch et al. 2008). A condensation temperature as  high as 1800K is suggested in the helium nova V445 Puppis (Ashok and Banerjee, 2003). V445 Puppis is of course an exceptional nova; it was the slowest among the dust-forming novae as can be seen from its isolated position in the Williams's et al. (2013) speed versus dust formation timescale plot. The dust shell, which formed initially in early 2001,  is extremely dense with the object continuing  to remain cocooned in dust even today as we find from regular monitoring by us from Mt Abu (and also from pvt. communication from Dr. Patrick Woudt).
\par
The grain size can be calculated by equating the  fraction of the outburst luminosity intercepted by the grains and which is subsequently re-emitted in the IR which is observationally a known quantity. This leads to the following relation for the grain radius (Gehrz et al. 1980, Evans et al. 2005, Gehrz et al. 2017):

\begin{equation}
a = \frac{L_O}{16 \pi A R^2 \sigma T^{(\beta +4)}}
\end{equation}

where $R$ is the nova-to-grain distance at a time $t$  assumed to be given by $v{\times}t$ where $v$ is the velocity,  $L_O$ is the outburst luminosity and $\sigma$ is the Stefan-Boltzmann constant. $L_O$  is assumed to remain constant, as would be expected for a CN during the evolutionary phase discussed here. A value of $v$ = 650 km/s is adopted using the HWHM of the  H$\alpha$ profile of Figure 6 in section 3.2. The computed grain sizes for AC and GR grains are listed in Table 6 and also shown in Figure 11. From  Figure 11 (top panel) a large increase by a factor of 6  is seen  in dust mass between the first and third  epochs (+19.21d to 36.18d)  corresponding to considerable growth in the number of grains subsequent to grain nucleation. Keeping in mind that the mass estimates of day +19.2 are upper estimates, the increase in mass during this period is expected to  be more than six-fold. This growth period is followed by  a phase where  the mass remains fairly constant over the remaining course of the observations  indicating neither significant growth nor destruction of the grains.
\par
Apart from sputtering and chemi-sputtering processes which can cause grain destruction (Evans et al. 2016 and references therein),  X-ray emission from the hot central WD can  be responsible for dust destruction principally by  grain heating and grain charging (Fruchter et al. 2001). Grain charging, which can lead to electrostatic stresses building up within a grain that is greater than the grain's tensile strength, is often the more important mechanism for destruction (Fruchter et al. 2001). Nova V339 Del, for e.g.,   passed through a long super soft phase (SS) and the  X-ray radiation, as shown by Evans et al. (2016),  could have been responsible for dust destruction. However in the case of the present nova, there were no reports of the detection of any X-ray emission from it.  Even if it did pass through a X-ray emission phase the possibility still remains that the grains could have been shielded from destruction by residing in dense clumps or regions of high density (this is discussed further below) into which the hard radiation  could not penetrate fully. However, observations taken lately  on 12 October 2016 using deep exposures failed to detect the nova up to limiting magnitudes of $J$ = 16.5 and $K$ = 14.75. It is possible that  the dust has been destroyed by this stage or its  emission dropped below detection limits. Alternatively it still could have been present but the dust SED may have shifted to longer wavelengths.
\par
The middle panel of Figure 11 shows the cooling of the dust  with time which is well approximated by a linear relation $T$ = 1619.6289 - 8.2699$\times$$t$ where $t$ is the time (this is a least squares minimized fit). This fit, which excluded the first data point,  was used to extrapolate and estimate the dust temperature of 1461 $\pm$ 15 on day 19.2 (the first data point) which is listed in Table 6 and which was used to estimate the dust mass on that day.
\par
The bottom panel shows the evolution of the grain radius with time. Surprisingly the grain radius shows an initial decrease followed by an almost constant or a mildly increasing trend. The latter behavior implies that the destructive processes that could reduce grain size are absent e.g. grain sputtering or grain shattering due to hard radiation. A similar inference was reached earlier from the evolution of the dust mass. The decrease in radius is hard to explain because grains are expected to grow in size after the initial nucleation phase. A plausible reason for this decrease could lie in the  simple assumptions behind equation 8 used to determine grain radius. A spherical geometry has been assumed and  the dust shell radius  is assumed to evolve linearly with time following $R$   = $v{\times}t$. However, high resolution spatial observations using  either  adaptive optics or interferometry show that the bipolar shape is fairly common in novae. In fact, time may show - as has been the case for  planetary nebulae (PNe) - that the bipolar shape may be the  generalized shape  that most CNe possess during the early expansion stages.  Examples of bipolar morphology at the early stages are clearly established either through direct  imaging or interferometry in V445 Puppis (Woudt et al. 2009), RS Oph (O'Brien et al. 2006), V1280 Sco (Chesneau et al. 2012), Nova Mon 2012 (Chomiuk et al. 2014) and V339 Del (Schaefer et al. 2014). There are old nova remnants too  like HR Del that have retained the bipolar morphology long after the original outburst.
\par
A tacit expectation of sphericial morphology was perhaps the  norm for PNe till the interacting winds scenario,  proposed by Kwok et al. (1978) and  subsequent elaborate and rigorous models developed by other workers (Balick $\&$ Frank 2002 and references therein), showed that the hourglass (or equivalently the bipolar shape) could be easily generated in the interacting winds scenario with the spherical morphology being just one special sub-class. The generic requirement to generate such a bipolar morphology was a density enhancement at the equatorial region (for e.g. caused by  matter concentrated in the equatorial regions that is left over from the common envelope phase) that restricts the wind  (or nova ejecta) from expanding freely  in the equatorial direction but allows an easy flow in the polar direction. The matter in the poles is then expected to  expand very rapidly while that in the equatorial waist is expected to   expand very slowly thereby creating  different expansion velocities across the bipolar structure. An example of this may be seen from the  kinematics of the ejecta in the nova  V455 Pup (Fig 6 , Woudt et al. 2009) wherein the equatorial matter at the waist barely expands with time. The nova under study here may also have similar  kinematical behavior. The double peaked velocity profile in Figure 6  indicates that there is a departure from spherical symmetry in the ejecta flow and a bipolar flow cannot be ruled out. Thus this nova may have an equatorial density enhancement which additionally  may be the site where the bulk of the dust resides. This is quite likely  because such  regions, by virtue of being dense,  are likely to have the critical density (Gehrz 1988) needed for grain nucleation  to occur while also  being  shielded from hard radiation that could potentially cause dust destruction. If such circumstances exist in this nova, then  equation 8  suggests that it is difficult to assign  precise meaning or values to the terms $R$ or $v$.
\par
There are  other complexities that affect a precise determination of the kinematics. First,  the velocity of the flow could change significantly between epochs. Observations which probed the nova's spatial development at very high resolution were not available earlier. But now, this is clearly seen from interferometric observations of  V339 Del which show that  there are considerable changes in the expansion rate with time (Schaefer et al. 2014). Specially relevant during the dust formation stage is the sharp drop in velocity immediately  after dust onset followed immediately  thereafter  by a rapid increase in the expansion rate (Figure 2, Schaefer et al. 2014). The second complexity is in the varying locations of the dust sites (hence giving rise to different $R$ values). For e.g., apart from showing that dust emission emanates from all across  the nova, V1280 Sco showed preferential and enhanced dust concentrations in the polar caps. All these factors above render it difficult to  assign a unique definition to the radius  $R$ in equation 8 or to justify the simple assumption that $R$ = v$\times$t as we have used here. We feel these simple assumptions can lead to imprecision in the results and the decrease in grain radius seen in the initial two epochs could be a consequence of this. However, it is felt that the principal result from this analysis  is  that the grain radii determined here for both amorphous and graphitic grains, are typically of the order of a micron or slightly more. This is similar to what is found in other novae when using the same formalism that has been adopted here (Gehrz et al. 1980, Gehrz 1988, Evans et al. 2016).

\begin{figure}
\centering
\includegraphics[bb= 133 2 442 412, width= 3.25in,clip]{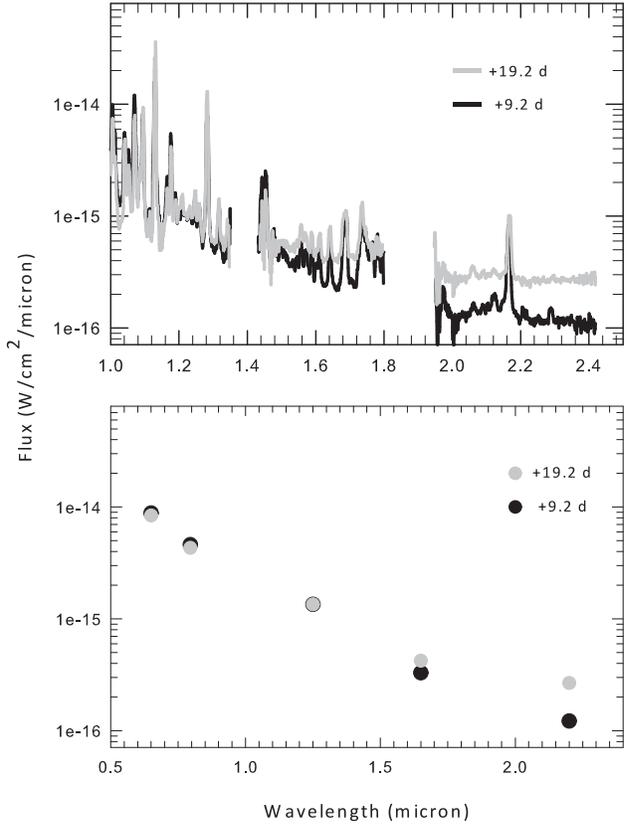}
\caption[]{ Top panel: changes seen in the spectra between days 9.2 and 19.2 showing the development in the latter epoch of an infrared excess  in the $H$ and $K$ bands due to dust. Bottom panel similarly shows the IR excess developing on day 19.2 by using R$_{c}$,I$_{c}$,$J,H$,$K_{s}$  broad band photometry. In both panels, the spectra and photometric fluxes  of days 9.2 and  19.2 have been scaled so as to match each other in the  $J$ band.  More details are given in the text.}
\label{fig-EarliestDust}
\end{figure}

\begin{table*}
\centering
\caption{Dust parameters and their evolution.}
\begin{tabular}{ccccccccccc}
\hline
$t$      & $(\lambda$${f_{\lambda})}_{max}$ &           $T_{dust}$   &    &  Mass ($10^{-8}$$M_{\odot}$)&&gr. rad $a$ ($\mu$m)&gr. rad $a$ ($\mu$m)        \\
(days)   & ($10^{-12}$W$m^{-2}$)      &  $(K)$   & AC$^a$  &GR  &BB&AC&GR       \\

\hline
\hline
19.21$^b$ 	&	3.49	$\pm$	0.34	&	1461	$\pm$	15	&	0.22	$\pm$	0.03	&	0.34	$\pm$	0.05	&	0.32	$\pm$	0.04	 &	3.89	$\pm$	0.93	&	5.83	$\pm$	0.95	\\
28.17	&	8.35	$\pm$	0.11	&	1405	$\pm$	27	&	0.63	$\pm$	0.07	&	0.96	$\pm$	0.11	&	0.86	$\pm$	0.08	&	 2.12	$\pm$	0.60	&	3.23	$\pm$	0.62	\\
36.18	&	13.10	$\pm$	0.29	&	1308	$\pm$	14	&	1.38	$\pm$	0.10	&	2.20	$\pm$	0.17	&	1.80	$\pm$	0.12	&	 1.80	$\pm$	0.43	&	2.87	$\pm$	0.45	\\
37.10	&	10.12	$\pm$	0.15	&	1293	$\pm$	23	&	1.13	$\pm$	0.11	&	1.81	$\pm$	0.20	&	1.46	$\pm$	0.13	&	 1.81	$\pm$	0.50	&	2.90	$\pm$	0.52	\\
58.11	&	5.23	$\pm$	0.07	&	1146	$\pm$	42	&	1.03	$\pm$	0.19	&	1.77	$\pm$	0.37	&	1.22	$\pm$	0.20	&	 1.31	$\pm$	0.48	&	2.24	$\pm$	0.50	\\
59.12	&	7.62	$\pm$	0.24	&	1143	$\pm$	38	&	1.53	$\pm$	0.29	&	2.62	$\pm$	0.54	&	1.80	$\pm$	0.29	&	 1.28	$\pm$	0.45	&	2.20	$\pm$	0.47	\\
64.07	&	4.39	$\pm$	0.20	&	1092	$\pm$	31	&	1.09	$\pm$	0.20	&	1.92	$\pm$	0.38	&	1.24	$\pm$	0.20	&	 1.36	$\pm$	0.44	&	2.38	$\pm$	0.46	\\
80.06	&	2.64	$\pm$	0.03	&	950	$\pm$	12	&	1.27	$\pm$	0.09	&	2.42	$\pm$	0.19	&	1.30	$\pm$	0.08	&	 1.68	$\pm$	0.42	&	3.20	$\pm$	0.43	\\
\hline
\hline
\end{tabular}
\label{table-DustPara}
\begin{list}{}{}
 \item a : AC = amorphous carbon, GR = graphite, BB = blackbody, gr. rad = grain radius
  \item b : $(\lambda$${f_{\lambda})}_{max}$ and mass estimates for this epoch are upper limits.
\end{list}
\end{table*}


\section{Summary}

Multi-epoch near infrared  and optical photometric  and spectroscopic observations  are presented for  nova V1831 Aquilae (Nova Aquilae 2015) discovered in outburst on 2015 September 09. We demonstrate that NIR spectroscopy is self-contained by itself to establish whether a nova is of the Fe II or the He/N class; in this particular case V1831 Aql is shown to be of the Fe II class. The distance and extinction $A_{\rm v}$ to the nova are estimated to be 6.1 $\pm$ 0.5 kpc and $\sim$ 9.02 magnitudes respectively. From a Case B analysis of the optically thick hydrogen Brackett lines the lower limits on the electron density $n_e$  and ejecta mass are found to be  in the range $\sim$ $5\times 10^9$ cm$^{-3}$ to  $1.2 \times 10^{10}$ cm$^{-3}$  and $(2.4 - 5.7)\times 10^{-6}$  M$_{\odot}$ respectively. From [OI] line ratios we estimate the neutral oxygen temperature and mass to be $T_{e} = 5156 \pm 170K$ and $M_{OI}$ = $(8.8 \pm 2.2)$ $\times$ $10^{-5}$ $M_{\odot}$ respectively.  A rapid, approximately 10-15 fold increase is seen in the (HeI 1.083)/(Pa$\gamma$ 1.094) line ratio during the early stages of evolution. This possibly arises  due to the He I line transitioning from an optically thick to thin state as the ejecta expands and dilutes. Dust formation is found to   occur in V1831 Aql around 19.2 days after discovery at a condensation temperature of 1461 $\pm$ 15K following which the dust gradually cools down to 950K over a period of $\sim$ 50  days. Based on the condensation temperature we suggest a carbonaceous composition for the dust. The dust mass shows a significant increase following nucleation by more than a factor of six between +19.21d to 36.18d, which is best interpreted as being due to  considerable growth in the number of grains. This growth period is followed by  a phase where  the mass remains fairly constant over the remaining course of the observations  indicating neither significant growth nor destruction of the grains. Following nucleation, the grains instead of showing growth show a decrease in radii during the initial stages. The plausible reasons for this behavior, and the subsequent evolution of the grain size  are discussed.


\section{Acknowledgments}
The research work at the Physical Research Laboratory is funded by the Department of Space, Government of India. We would like to thank A.Frigo, P.Ochner, S.Dallaporta and U. Sollecchia for their assistance with some of the optical observations. We acknowledge the use of data from the AAVSO database.


\begin{thebibliography}{99}

\bibitem[]{b2} Almog, Y; Netzer, H, 1989, MNRAS, 238, 57
\bibitem[]{b2} Balick, B.,  Frank, A. 2002,  ARAA 40, 439
\bibitem[]{b2} Ashok, N. M.; Banerjee, D. P. K., 2003, A$\&$A 409, 1007
\bibitem[]{b2} Ashok, N. M.; Srivastava, M.K.; Banerjee, D. P. K., 2015,  Astron. Telegram, 8142, 1
\bibitem[\protect\citeauthoryear{Banerjee}{2012}]{b5} Banerjee D. P. K., Ashok N. M., 2002, A$\&$A, 395, 161
\bibitem[]{b2} Banerjee, D. P. K.; Ashok, N. M.; Venkataraman, V., 2012 Astron.Telegram, 4542,1
\bibitem[\protect\citeauthoryear{Banerjee}{2012}]{b5} Banerjee D. P. K., Ashok N. M., 2012, BASI, 40, 243
\bibitem[\protect\citeauthoryear{Banerjee}{2016}]{b5} Banerjee D. P. K., Srivastava, Mudit K., Ashok, N. M., Venkataraman, V.,2016,MNRAS, 455, 1, L109
\bibitem[]{b2} Benjamin, R. A., Skillman, E. D., \& Smits, D. P.1999, ApJ, 514, 307
\bibitem[]{b2}Blanco, A., Falcicchia, G, Merico, F., Ap$\&$SS, 1983, 89, 1638
\bibitem[]{b2} Cardelli, J.A., Clayton, G.C., Mathis, J. 1989, ApJ,  345, 245
\bibitem[]{b2} Chesneau, O., Banerjee, D.P.K.., Millour, F. et al 2008, A\&A, 487, 223
\bibitem[]{b2} Chesneau, O.,  et al. 2012, A\&A, 545, 63
\bibitem[]{b2} Chomiuk, L., et al., 2014, Nature, 514, 339
\bibitem[\protect\citeauthoryear{Das}{2008}]{b15} Das R. K., Banerjee D. P. K., Ashok N. M., Chesneau O., 2008, MNRAS, 391, 1874
\bibitem[\protect\citeauthoryear{Das}{2009}]{b7} Das R.K., Banerjee D.P.K., Ashok N.M.,  2009, MNRAS, 398, 375
\bibitem[\protect\citeauthoryear{Das}{2009}]{b7} Das R.K., Banerjee D.P.K., Ashok N.M., Soumen Mondal,  Bull. Astr. Soc. India, 2013, 41, 195
\bibitem[\protect\citeauthoryear{della Valle}{1995}]{b8} della Valle M., Livio M., 1995, ApJ, 452, 704
\bibitem[\protect\citeauthoryear{della Valle}{2002}]{b8} Della Valle et al.  2002, A\&A, 390, 155
\bibitem[\protect\citeauthoryear{Downes}{2000}]{b10} Downes R. A., Duerbeck H. W., 2000, AJ, 120, 2007
\bibitem[]{b2} Ederoclite, A., 2013, Highlights of Spanish Astrophysics VII, Proceedings of the X Scientific Meeting of the Spanish Astronomical Society (SEA), held in Valencia, July 9 - 13, 2012, Eds.: J.C. Guirado, L.M. Lara, V. Quilis, and J. Gorgas., pp.539-542).
\bibitem[]{b2} Evans A., Geballe T.~R., Rawlings J.~M.~C., Scott A.~D., 1996, MNRAS, 282, 1049
\bibitem[]{b2} Evans A., Tyne V. H., Smith O., Geballe T. R., Rawlings J. M. C., Eyres S.
P. S., 2005, MNRAS, 360, 1483
\bibitem[]{b2} Evans, A., Banerjee, D.P.K., Gehrz, R. D., Joshi, V., et al.,  2016, MNRAS, submitted.
\bibitem[]{b2} Fruchter A., Krolik J. H., Rhoads J. E., 2001, ApJ, 563, 597
\bibitem[]{b2} Fujii M., 2015, IAU Circ., 9278, 1
\bibitem[]{b2} Gehrz R. D., Hackwell J. A., Grasdalen G. I., Ney E. P., Neugebauer G., Sellgren K., 1980, ApJ, 239, 570
\bibitem[]{b2} Gehrz R.D., 1988, ARAA, 26, 377
\bibitem[]{b2} Gehrz R.D., Nye E.P., 1992, Icarus, 100, 162
\bibitem[]{b2} Gehrz, R. D., Evans, A., Helton, L. A., Shenoy, D. P., Banerjee, D.P.K., Woodward, C.E., Vacca, W.D.,
Dykhoff, D.A. et al.,  2015, 812, 132
\bibitem[]{b2} Gehrz, R. D., Evans, A.,  Woodward, C.E., Helton, L. A.,  Banerjee, D.P.K., Srivastava, M., Ashok, N.M., Joshi, V. et al., 2017, ApJ, under revision.
\bibitem[]{b2} Goranskij V. P.,Barsukova, E. A., 2015, Astron. Telegram, 8150, 1
\bibitem[]{b2} Joshi V., Banerjee D.P.K., Ashok N.M., 2014, MNRAS,443,559J
\bibitem[]{b2} Harrison, T. E., Stringfellow, G. S. 1994, ApJ, 437, 827
\bibitem[\protect\citeauthoryear{Hummer}{1987}]{b19} Hummer D. G., Storey P. J., 1987, MNRAS, 224, 801
\bibitem[]{b2} Kwok, S.; Purton, C. R.; Fitzgerald, P. M., 1978, ApJ, 219, L125
\bibitem[]{b2}Landolt A. U.,2009, AJ,137,5,4186
\bibitem[]{b2}Lodders, K.; Fegley, B., Jr., 1999, Asymptotic Giant Branch Stars, IAU Symposium 191, Eds: T. Le Bertre, A. Lebre, and C. Waelkens.
p. 279
\bibitem[]{b2}Lodders, K.; Fegley, B., Jr., 1995, Meteoritics, 30, 661
\bibitem[]{b2} Lynch D. K., et al., 2008, AJ, 136, 1815
\bibitem[]{b2}Maehara H., Fujii M., 2015, Astron. Telegram, 8127, 1
\bibitem[]{b2}Maccarone, T. J., 2015, Astron. Telegram, 8136, 1
\bibitem[]{b2}McLaughlin, D. B. 1960, in Stellar Atmospheres, ed. J. L. Greenstein (Chicago: Univ. of
Chicago Press)
\bibitem[\protect\citeauthoryear{Marshall}{2006}]{b25}  Marshall D. J., Robin A. C., Reyle C., Schultheis M., Picaud S., 2006, A\&A, 453, 635
\bibitem[]{b2}Munari, U., Siviero, A., Henden, A., Cardarelli, G., Cherini, G., Dallaporta, S., Dalla Via, G., Frigo, A., Jurdana-Sepic, R., Moretti, S., et al. 2008, A$\&$A, 492, 145
\bibitem[]{b2}Munari, U. Moretti, S., 2012, Baltic Astronomy, 21, 22
\bibitem[]{b2} Munari, U., Bacci, S., Baldinelli, et al.  2012, Baltic Astronomy, 21, 13
\bibitem[]{b2} Munari, U., Maitan, A., Moretti, S.,  Tomaselli, S.,  2015, New Astronomy, 40, 28
\bibitem[]{b2} Nakano S., 2015, IAU Circ., 9278, 1
\bibitem[]{b2} Naik, S,  Banerjee, D. P. K.; Ashok, N. M.; Das, R. K., 2010, MNRAS, 404, 367
\bibitem[]{b2}O'Brien, T. J.; Bode, M. F.; Porcas, R. W.; Muxlow, T. W. B.; Eyres, S. P. S.; Beswick, R. J.; Garrington, S. T.; Davis, R. J.; Evans, A., Nature, 2006, 442, 279
\bibitem[]{b2} Payne-Gaposchkin C., 1957, The Galactic Novae, North-Holland Publishing Co., Amsterdam
\bibitem[]{b2} Pontefract, M; Rawlings,J.M.C., 2004, MNRAS, 347, 1294
\bibitem[]{b2} Porter R.L., Bauman R. P., Ferland G. J., and MacAdam K. B., 2005,  ApJ, 622, L73
\bibitem[\protect\citeauthoryear{Rieke}{1985}]{b38} Rieke G.H., Levofsky M.J.,1985, ApJ, 368, 468
\bibitem[]{b2}Schaefer G. H., et al. 2014, Nature, 515, 234
\bibitem[]{b2}Shappee B.J. et al., 2015, Astron. Telegram, 8126, 1
\bibitem[]{b2}Shafter, A. W.; Darnley, M. J.; Hornoch, K.; Filippenko, A. V.; Bode, M. F.; Ciardullo, R.; Misselt, K. A.; Hounsell, R. A.; Chornock, R.; Matheson, T., 2011, ApJ,734, 12
\bibitem[\protect\citeauthoryear{Schlafly}{2011}]{b39} Schlafly E., Finkbeiner D.P., 2011, ApJ, 737, 103
\bibitem[]{b2} Speck, A.K., Barlow, M.J., Sylvester, R.J., Hofmeister, A.M, 2000, Astron. Astrophys. Suppl. Ser. 146, 437
\bibitem[]{b2}Srivastava, M. K.,  Banerjee, D. P. K.,  Ashok, N. M., 2015,  Astron. Telegram, 8332, 1
\bibitem[\protect\citeauthoryear{Srivastava}{2015}]{b41} Srivastava M., Ashok N. M., Banerjee D. P. K., Sand, D., 2015, MNRAS, 454, 1297
\bibitem[]{b2} Stanek, K.Z.,  1996, ApJ, 460, L37
\bibitem[\protect\citeauthoryear{Storey}{1995}]{b43} Storey P. J., Hummer D. G., 1995, MNRAS, 272, 41
\bibitem[\protect\citeauthoryear{Strope}{2010}]{b36} Strope R. J., Schaefer B. E.,  Henden A. A., 2010, AJ, 140, 34
\bibitem[]{b2} Tanaka J., Nogami D., Fujii M., Ayani K., Kato T., Maehara H., Kiyota S., Nakajima K., 2011, PASJ, 63, 911
\bibitem[]{b2} Todini, P.; Ferrara, A., 2001, MNRAS, 325, 726
\bibitem[\protect\citeauthoryear{vandenBergh}{1987}]{b45} van den Bergh S., Younger P.S., 1987, A\&AS, 70, 125
\bibitem[]{b2} Warner, B. 1995, in Cataclysmic Variable Stars, (Cambridge: Cambridge Univ. Press)
\bibitem[]{b2} Williams P.M., Longmore A. J., Geballe T. R., 1996, MNRAS, 279, 804
\bibitem[]{b2} Williams R. E., Hamuy, M., Phillips, M. M., Heathcote, S. R., Wells, L., Navvarete, M., 1991, ApJ, 376, 721
\bibitem[]{b2} Williams R. E., 1992, AJ, 104, 725
\bibitem[]{b2} Williams R.E., 1994, ApJ, 426, 279
\bibitem[\protect\citeauthoryear{Williams}{2012}]{b47} Williams, R.E., 2012, AJ, 144, 98
\bibitem[]{b2} Williams, S. C.; Bode, M. F.; Darnley, M. J.;Evans, A.;  Zubko, V., Shafter, A.W.,  2013, ApJ,  777, L32
\bibitem[]{b2} Woudt, P. A., Steeghs, D., Karovska, M., et al. 2009, ApJ, 706, 738








\end{thebibliography}
\end{document}